\documentclass[12pt]{article}
\usepackage{amsmath,amssymb,epsfig}
\makeatletter \@addtoreset{equation}{section} \makeatother
\renewcommand{\theequation}{\thesection.\arabic{equation}}
\newcommand{\ba}{\begin{array}}
\newcommand{\ea}{\end{array}}
\newcommand{\beq}{\begin{equation}}
\newcommand{\eeq}{\end{equation}}
\newcommand{\bea}{\begin{eqnarray}}
\newcommand{\eea}{\end{eqnarray}}

\def\bce{\begin{center}}
\def\ece{\end{center}}

\def\nonu{\nonumber}

\def\be{\beta}

\newcommand{\tr}{\mbox{Tr}}

\def\eps6{{\displaystyle \mathop{\epsilon}^{6}}{}}

\def\nab6{{\displaystyle \mathop{\nabla}^{6}}{}}

\def\0{{\sst{(0)}}}
\def\1{{\sst{(1)}}}
\def\2{{\sst{(2)}}}
\def\3{{\sst{(3)}}}
\def\4{{\sst{(4)}}}
\def\5{{\sst{(5)}}}
\def\6{{\sst{(6)}}}
\def\7{{\sst{(7)}}}
\def\8{{\sst{(8)}}}

\def\ba{\begin{array}}
\def\ea{\end{array}}
\def\beq{\begin{equation}}
\def\eeq{\end{equation}}
\def\be{\begin{equation}}
\def\ee{\end{equation}}

\def\tr{\mathop{\rm tr}}

\def\eps{\epsilon}

\def\ba{\begin{array}}
\def\ea{\end{array}}
\def\beq{\begin{equation}}
\def\eeq{\end{equation}}
\def\be{\begin{equation}}
\def\ee{\end{equation}}

\def\tr{\mathop{\rm tr}}

\def\eps{\epsilon}

\newcommand{\bean}{\begin{eqnarray*}}
\newcommand{\eean}{\end{eqnarray*}}

\begin{document}
\thispagestyle{empty} \addtocounter{page}{-1}
   \begin{flushright}
KIAS-P08015 \\
\end{flushright}

\vspace*{1.3cm}

\centerline{ \Large \bf  Meta-Stable Brane Configurations }
\vspace{.3cm} 
\centerline{ \Large \bf by Higher Order Polynomial Superpotential } 
\vspace*{1.5cm}
\centerline{{\bf Changhyun Ahn} 
} 
\vspace*{1.0cm} 
\centerline{\it 
Department of Physics, Kyungpook National University, Taegu
702-701, Korea} 
\vspace*{0.8cm} 
\centerline{\tt ahn@knu.ac.kr} 
\vskip2cm

\centerline{\bf Abstract}
\vspace*{0.5cm}

We construct the type IIA
nonsupersymmetric meta-stable brane
configuration consisting of $(2k+1)$ NS5-branes and D4-branes 
where the electric gauge theory superpotential has 
an order $(2k+2)$ polynomial for the bifundamentals.
We find a rich pattern of nonsupersymmetric meta-stable states as well
as the supersymmetric stable ones.
By adding the orientifold 4-plane to this brane
configuration,
we also describe the intersecting brane configuration of type IIA 
string theory corresponding to the meta-stable nonsupersymmetric 
vacua of corresponding gauge theory.

\baselineskip=18pt
\newpage
\renewcommand{\theequation}
{\arabic{section}\mbox{.}\arabic{equation}}

\section{Introduction}

The dynamical supersymmetry breaking in meta-stable vacua \cite{ISS,IS} 
occurs
in the ${\cal N}=1$ gauge theory with massive fundamental 
flavors.
The mass term
for the quarks in the electric superpotential has led to the fact that
some of the F-term equations of magnetic superpotential
cannot  be satisfied and the
supersymmetry is broken.
The meta-stable brane constructions of type IIA string theory 
have been studied in \cite{OO1,FGU,BGHSS}.  

In \cite{GK0710-1,GK0710}, other kind
of 
the type IIA nonsupersymmetric meta-stable 
brane configuration was constructed by considering  
the additional  quartic
term for the quarks in the electric superpotential besides the mass term.
This extra deformation in the supersymmetric gauge theory 
corresponds to the rotation of D6-branes 
along the (45)-(89) directions in type IIA string theory. 
Since the extra quartic term gives rise to the fact that all
the F-term equations are satisfied, 
only
supersymmetric ground states are present classically.
The nonsupersymmetric ground states arise only after 
the gravitational attraction of NS5-brane is considered.

By adding the orientifold 6-plane to this brane
configuration \cite{GK0710-1},
the meta-stable nonsupersymmetric 
brane configuration corresponding to 
the supersymmetric gauge theory with symmetric flavor
as well as fundamental flavors was found \cite{Ahn07-11}. 
For the antisymmetric flavor plus fundamental flavors case,
the corresponding meta-stable brane configuration was also 
described in \cite{Ahn08-1}.
Moreover, the meta-stable brane
configuration consisting of three NS5-branes, D4-branes and 
anti-D4-branes was constructed in \cite{Ahn08-1two} by deforming the
theory described in \cite{GK}.

As suggested in \cite{GK0710-1},
what happens when ``multiple'' NS-branes  are rotated relatively? 
One expects, in the gauge theory side, 
that the higher order term for the bifundamentals besides the mass term 
appears in the superpotential. 
Since there exist multiple outer NS-branes,
a rich pattern of nonsupersymmetric meta-stable states  
is possible because the flavor D4-branes can suspend between 
these two multiple NS-branes in many different way.
We'll elaborate the meta-stable vacuum structure both in the gauge
theory side and the type IIA string theory side.

First, we construct the meta-stable brane configuration 
by rotating and displacing these multiple NS-branes relatively,  
taking the Seiberg dual, and splitting, reconnecting or displacing 
the flavor and color D4-branes.
When these multiple NS-branes are coincident, then 
the behavior for the meta-stable brane construction looks similar to 
the single NS-brane case 
\cite{Ahn08-1two}. However, when they are not coincident with 
each other, many different pattern for the meta-stable 
brane construction occurs.  
Secondly, we describe the meta-stable brane configuration 
by adding an orientifold 4-plane to 
this brane configuration and following the previous 
method done in unitary gauge group, analyze 
the vacuum structure for the different gauge theory with matters,
and present the different aspects 
in the presence of orientifold 4-plane.

In section 2, we review the type IIA brane configuration corresponding
to the ${\cal N}=1$ $SU(N_c) \times
SU(N_c')$ gauge theory 
with the bifundamentals and deform this theory 
by adding both the mass term
and the higher order term for the bifundamentals. 
Then we construct the dual 
${\cal N}=1$ $SU(\widetilde{N}_c) \times SU(N_c')$ gauge 
theory with corresponding dual
matter as well as gauge singlet. 
We describe the nonsupersymmetric meta-stable
minimum  and present 
the corresponding 
intersecting brane configuration of type IIA string
theory.

In section 3, we review the type IIA brane configuration corresponding
to the ${\cal N}=1$ $Sp(N_c) \times
SO(2N_c')$ 
gauge theory 
with a bifundamental and deform this theory by adding the mass term
and the higher order term for the bifundamental. 
Then we describe the dual  
${\cal N}=1$ $Sp(\widetilde{N}_c) \times SO(2N_c')$ gauge 
theory with corresponding dual matters. 
We construct the nonsupersymmetric meta-stable
minimum  and present 
the corresponding intersecting brane configuration of type IIA string
theory. 

In section 4,  we make some comments for the future directions.  

\section{Meta-stable brane configuration with $(2k+1)$ NS-branes }

\subsection{Electric theory}

The type IIA brane configuration for an  
${\cal N}=1$ supersymmetric gauge theory with
gauge group
$
SU(N_c) \times SU(N_c')
$
and 
a bifundamental $X$ in the representation 
$({\bf N_c, \overline{N_c'}})$ and its conjugate field $\widetilde{X}$ 
in the representation $({\bf \overline{N_c}, N_c'})$ 
can be constructed as follows \cite{BH,AT97}: 
the middle NS5-brane(012345),
the left $k$ $NS5_L'$-branes(012389), the right $k$ 
$NS5_R'$-branes(012389), $N_c$-  and $N_c'$-color 
D4-branes(01236) where the number of $NS5_{L,R}'$-branes 
is a positive integer $k \geq 1$.
The $N_c$-color D4-branes 
are suspending between the 
middle NS5-brane whose $x^6$ coordinate is $x^6=0$ and 
the $NS5_R'$-branes while 
the $N_c'$-color D4-branes 
are suspended between the $NS5_L'$-branes and the
middle NS5-brane.
We consider the arbitrary numbers of color D4-branes 
with the constraint 
$N_c' \geq N_c$.

Let us deform 
this theory which has vanishing superpotential. 
Displacing the two kinds of $NS5_{L,R}'$-branes relative each other in the 
$v$
direction corresponds to turning on a quadratic
superpotential
for the bifundamentals $X$ and $\widetilde{X}$.
Rotating these $NS5_{L,R}'$-branes in the 
$(v,w)$ plane
corresponds to turning on a quartic($k=1$) or higher order($k\geq 2$) 
superpotential
for the bifundamentals $X$ and $\widetilde{X}$ \cite{BH,BHKL}
\footnote{
We introduce two complex coordinates
$ v \equiv x^4 + i x^5$ and $w \equiv x^8 + i x^9$ \cite{GK98}.}.
Then the deformed electric superpotential, 
by changing the position of the left $k$
$NS5_L'$-branes only, 
can be written as 
\cite{BH,BHKL,ILS}
\bea
W_{elec} = -\frac{\alpha}{2}   
\tr (X \widetilde{X})^{k+1} +  \tr m X \widetilde{X},
\qquad
\alpha = \frac{\tan \theta}{\Lambda}, 
\qquad m = \frac{v_{NS5_{-\theta}}}{2\pi \ell_s^2}. 
\label{superelectric}
\eea
The $k$ $NS5_L'$-branes are moving to the $+v$ direction together with $N_c'$
D4-branes and are rotating 
by an angle $-\theta$ in $(w,v)$-plane.
Then the 
$v$ coordinate of $NS5_L'$-branes is denoted by $v=+
v_{NS5_{-\theta}}$ and we denote the $NS5_L'$-branes by the
$NS5_{-\theta}$-branes after deformation. 
Definitely, the number of multiple NS-branes $k$ should be less than 
$N_c'$ because we'll see, in the magnetic brane configuration, that
the 
$N_c'$-flavor D4-branes are suspended between two $k$ NS-branes.
We draw the electric brane
configuration in Figure 1.

\begin{figure}[ht]
   \epsfxsize=2.5in 
\centerline{\epsffile{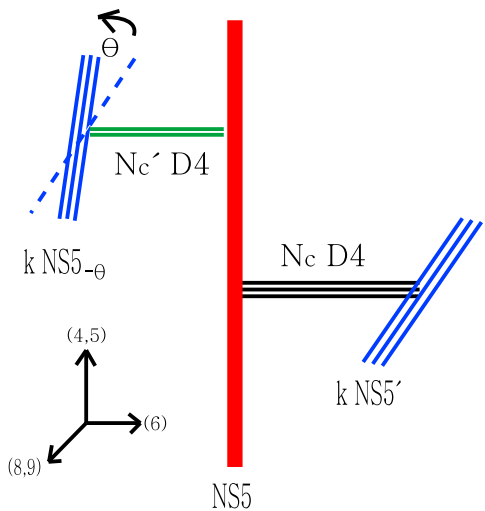}}
   \caption[FIG. \arabic{figure}.]{ 
The  ${\cal N}=1$ supersymmetric 
electric brane configuration for the gauge group $SU(N_c) \times
SU(N_c')$ and the bifundamentals $X$ and $\widetilde{X}$  
with  
nonvanishing mass and higher order terms
for the bifundamentals. 
Note that there are multiple outer NS-branes.
A 
``rotation'' of $k$ $NS5_L'$-branes in $(w,v)$-plane, which become
$NS5_{-\theta}$-branes, 
corresponds to 
a higher order term for the bifundamentals while 
a ``displacement'' of $k$ $NS5_L'$-branes in $+v$ direction corresponds to a
mass term for the bifundamentals.
}
\end{figure}

The solution for the supersymmetric vacua can be written as 
$X \widetilde{X} =\left[\frac{2m}{\alpha(k+1)}\right]^{\frac{1}{k}}$ 
through the F-term conditions.
This breaks 
the gauge group $SU(N_c) \times SU(N_c')$ 
to $SU(p_0)$, $SU(N_c'-N_c+p_0)$,
and $U(p_1) \times U(p_2) \times \cdots \times U(p_k)$ with
$\sum_{i=0}^k p_i =N_c$ \cite{ILS}.
The $p_0$ is the number of eigenvalues which are zero with 
$X=0=\widetilde{X}$.
When the middle NS5-brane moves to $+w$ direction, then the three
kinds of NS-branes intersect in three points in $(v,w)$-plane. 
Then we consider that $p_0$
D4-branes are connecting between the middle NS5-brane and the
$NS5_R'$-branes. One can decompose $N_c$ D4-branes as 
$p_0$ plus $(N_c-p_0)$ D4-branes.
The latter can be reconnected with the same number of D4-branes 
among $N_c'$ D4-branes.
Then the remaining $(N_c'-N_c+p_0)$
D4-branes are connecting between the $NS5_{-\theta}$-branes and 
the middle NS5-brane. 
Then $(N_c-p_0)$ D4-branes are suspended between $k$
$NS5_{-\theta}$-branes and $NS5_R'$-branes.
By separating these NS-branes transversely, each $p_i$
D4-branes(where $i=1, 2, \cdots, k$) 
are connecting between the $i$-th $NS5_{-\theta,i}$-brane and the
$i$-th $NS5_{R,i}'$-brane directly.

\subsection{Magnetic theory}

Let us apply the Seiberg dual to the 
$SU(N_c)$ factor only and 
the middle NS5-brane and the right
$NS5_R'$-branes are 
interchanged each other \cite{GK98}. Then the number of color $\widetilde{N}_c$
is given by $\widetilde{N}_c=N_c'-N_c$ connecting between the $NS5_R'$-branes
and the NS5-brane. One can easily check that the linking numbers of 
three kinds of NS-branes are preserved during this dual process.
For example, the linking number of $NS5_R'$-brane in an electric
theory is given by $l_e= -\frac{N_c}{k}$ while the one in magnetic
theory is given by $l_m=\frac{\widetilde{N}_c}{k}-\frac{N_c'}{k}$.
By introducing $N_c'$ D4-branes and $N_c'$ 
anti-D4-branes  between $NS5_R'$-branes and   
NS5-brane, reconnecting the former with  
the $N_c'$ D4-branes(that are connecting between the 
$NS5_{-\theta}$-branes and the $NS5_R'$-branes) and moving those 
combined D4-branes
to $+v$-direction, 
one gets the final Figure 2(with $l=0$) and there exist 
$(N_c'-\widetilde{N}_c)$ anti-D4-branes between $NS5_R'$-branes and   
NS5-brane when the distance between the multiple NS-branes 
is large \cite{Ahn08-1two}.

\begin{figure}[ht]
   \epsfxsize=2.5in 
\centerline{\epsffile{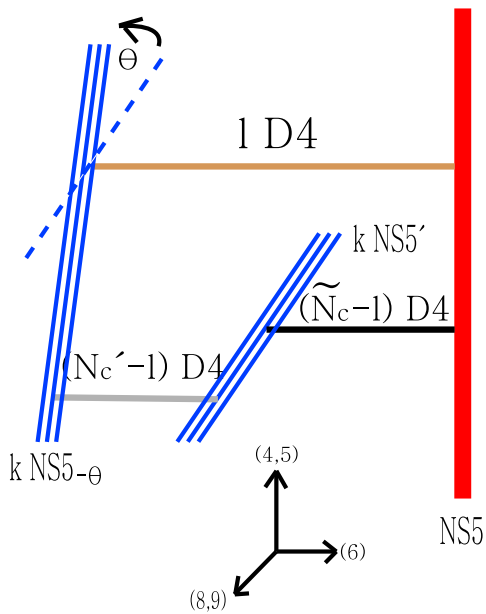}}
   \caption[FIG. \arabic{figure}.]{ 
The  ${\cal N}=1$ supersymmetric
magnetic brane configuration corresponding to Figure 1 with 
a misalignment between D4-branes when the gravitational potential of
the NS5-brane is ignored. 
The $N_c'$ flavor D4-branes connecting between
$NS5_{-\theta}$-branes and $NS5_R'$-branes are splitting into $(N_c'-l)$ and
$l$ D4-branes. 
The intersection between $NS5_{-\theta}$-branes and $(N_c'-l)$
D4-branes 
is given by $(v,w)=(0, +v_{NS5_{-\theta}} 
\cot \theta)$ while the one between  
$NS5_{-\theta}$-branes and $l$
D4-branes 
is given by $(v,w)=(+v_{NS5_{-\theta}}, 0)$. 
}
\end{figure}

The dual gauge group is given by
$
SU(\widetilde{N}_c=N_c'-N_c) \times SU(N_c')$ and
the matter contents are the field $Y$ in the representation 
$({\bf \widetilde{N}_c, \overline{N_c'}})$,
its complex conjugate field $\widetilde{Y}$ in the  
representation  $({\bf \overline{\widetilde{N}_c}, N_c'})$
and  the gauge singlet $M \equiv X \widetilde{X}$ 
in the representation  
$({\bf 1,{N_c^{\prime}}^2-1}) \oplus ({\bf 1, 1})$.
Then the dual magnetic superpotential, by adding the mass term
and the higher order term 
for the bifundamentals $X$ and $\widetilde{X}$ from an electric theory
(\ref{superelectric}) to the cubic
superpotential between the dual magnetic matters, is given by
\bea
W_{dual} =  \frac{1}{\Lambda} \tr M Y \widetilde{Y}- 
\frac{\alpha}{2} \tr M^{k+1}  + \tr m M. 
\label{superpo11}
\eea
Note that in the first term, the trace means that 
the $\widetilde{N}_c$ indices of $Y$ and $\widetilde{Y}$ are contracted each
other. The two $N_c'$ indices of $M$ are contracted with those of
$Y$ and $\widetilde{Y}$ respectively.
The trace in the second and third terms runs 
over the $N_c'$ indices
\footnote{As pointed out in \cite{Ahn08-1two}, the conditions
 $b_{SU(\widetilde{N}_c)}^{mag} 
< 0$ and $b_{SU(N_c)} > 0$ imply
that $N_c' < \frac{3}{2} N_c $. Then the range for the $N_c'$ 
can be written as $ N_c < N_c' 
< \frac{3}{2} N_c $. 
At the scale $\Lambda_1$, the $SU(N_c)$ theory is strongly
coupled and the Seiberg duality occurs. The coefficients of beta function 
$b_{SU(\widetilde{N}_c)}^{mag}$ becomes negative and 
$b_{SU(N_c')}^{mag}$
becomes positive. 
Then at energy scale lower than Landau pole $\Lambda_1$, 
the theory is weakly coupled. 
Then under the constraint, $\Lambda_2 <<  \left(
  \frac{\Lambda_1}{\mu}\right)^b
\Lambda_1 << \Lambda_1$
where $b \equiv 
\frac{b_{SU(\widetilde{N}_c)}^{mag}-b_{SU(N_c')}^{mag}}{b_{SU(N_c')}}$, 
one can ignore the contribution from the gauge coupling of 
$SU(N_c')^{mag}$ at the supersymmetry breaking scale. 
Then one can use the magnetic 
superpotential (\ref{superpo11}) safely.  
Then
a further generalization looks straightforward once the tuning of the 
parameters is chosen in this way. }. 
In order to obtain the supersymmetric vacua, 
one computes the F-term equations for 
the superpotential (\ref{superpo11}):
\bea
 M Y & = & 0, \qquad \widetilde{Y} M =0, 
\nonu \\
-\frac{1}{\Lambda} Y  \widetilde{Y} & = & m -\frac{\alpha(k+1)}{2} M^k. 
\label{Fterm}
\eea

Let us first consider all the 
$NS5_{-\theta}$-branes and $NS5_R'$-branes
are coincident with each other. Later, we'll split them 
in transverse direction.

$\bullet$ Coincident $NS5_{-\theta}$-branes and $NS5_R'$-branes

Since the eigenvalues for the meson field $M$
are either $0$ or $\left[\frac{2m}{\alpha(k+1)}\right]^{\frac{1}{k}}$, 
through F-term equations, one takes
$N_c' \times N_c'$ matrix $M$ with $l$'s eigenvalues $0$ and $(N_c'-l)$'s
eigenvalues $\left[\frac{2m}{\alpha(k+1)}\right]^{\frac{1}{k}}$ as follows:
\bea
M = \left(
\begin{array}{cc}
0{\bf 1}_l & 0  \\
0 & \left[\frac{2m}{\alpha(k+1)}\right]^{\frac{1}{k}} {\bf 1}_{N_c'-l}
\end{array}
\right)
\label{M0}
\eea
where $l=1, 2, \cdots, N_c'$.
In the brane configuration of Figure 2, the $l$ of the
$N_c'$-flavor D4-branes are connected with $l$ of $\widetilde{N}_c$-color
D4-branes
and the resulting $l$ D4-branes stretch from the $NS5_{-\theta}$-branes to
the NS5-brane directly 
and the intersection point between the 
$l$ D4-branes and the NS5-brane is given by 
$(v, w)=(+v_{NS5_{-\theta}}, 0)$.
This corresponds to  exactly the $l$'s eigenvalues $0$ of 
$M$ in (\ref{M0}).
Now the remaining $(N_c'-l)$-flavor D4-branes between 
the $NS5_{-\theta}$-branes and 
the $NS5_R'$-branes correspond to the remaining eigenvalues 
of $M$ in (\ref{M0}), i.e.,   
$\left[\frac{2m}{\alpha(k+1)}\right]^{\frac{1}{k}} {\bf 1}_{N_c'-l}$.
The intersection point between the 
$(N_c'-l)$ D4-branes and the $NS5_R'$-branes is given 
by $(v, w)=(0, +v_{NS5_{-\theta}} \cot \theta)$ from trigonometric 
geometry \cite{GK0710-1,Ahn08-1two}.

By substituting (\ref{M0}) into the last equation of (\ref{Fterm}), 
one obtains
the following expectation value 
\bea
Y  \widetilde{Y} = \left(
\begin{array}{cc}
-m \Lambda {\bf 1}_l & 0  \\
0 & 0 {\bf 1}_{N_c'-l}
\end{array}
\right).
\label{solqq}
\eea
In the $l$-th vacuum the gauge symmetry is broken to $SU(\widetilde{N}_c-l)$
and 
the supersymmetric vacuum drawn in Figure 2 with $l=0$ has 
$Y \widetilde{Y} =0$ and the gauge group 
$SU(\widetilde{N}_c)$ is unbroken. 
If we replace $k$ $NS5_{-\theta}$-branes with $N_f$ D6-branes
and  $k$ $NS5_{R}'$-branes with a single $NS5_R'$-brane with
$l=\widetilde{N}_c$, 
then 
the meta-stable brane configuration of Figure 2 reduces to the one in 
\cite{OO1,FGU,BGHSS,Ahn07-3,Ahn07-8}.

On the other hand,
the theory has many nonsupersymmetric meta-stable ground states
due to the fact that there exists an attractive
gravitational interaction
between the flavor D4-branes and the NS5-brane from the DBI action \cite{GK}.
When we rescale the meson field as
$M = h \Lambda \Phi $,
then the Kahler potential for $\Phi$ is canonical and the magnetic
quarks $Y$ and $\widetilde{Y}$ 
are canonical near the origin of field space.
Then the magnetic superpotential (\ref{superpo11}) 
can be rewritten in terms of $\Phi, Y$ and $\widetilde{Y}$
\bea
W_{dual} = 
 h \Phi  Y   \widetilde{Y} 
 +  
\frac{ \mu_{\phi}}{2} h^{k+1} \tr \Phi^{k+1}- h \mu^2 \tr \Phi 
\label{Dual}
\eea
with the new couplings
$
\mu^2 = -m \Lambda$ and  
$\mu_{\phi} = -\alpha \Lambda^{k+1}$.

Now one splits 
the $(N_c'-l) \times (N_c'-l)$
block  at the lower right corner of $M$ and $Y
\widetilde{Y}$ 
into blocks of 
size $n$ and $(N_c'-l-n)$ and then (\ref{M0}) and (\ref{solqq}) are 
rewritten as follows \cite{GK0710}:
\bea
h\Phi = \left(
\begin{array}{ccc}
0{\bf 1}_l & 0 & 0  \\
0 & h \Phi_n & 0 \\
0 & 0 & \left[\frac{2\mu^2}{\mu_{\phi}(k+1)}\right]^{\frac{1}{k}} 
{\bf 1}_{N_c'-l-n}
\end{array}
\right), \qquad
Y  \widetilde{Y} = \left(
\begin{array}{ccc}
\mu^2 {\bf 1}_l & 0 & 0  \\
0 & { \varphi}  \widetilde{\varphi}  &  0 \\
0 & 0 & 0{\bf 1}_{N_c'-l-n}
\end{array}
\right).
\label{Eigen}
\eea
Here $\varphi$ and $\widetilde{\varphi}$ are $n \times (\widetilde{N}_c-l)$
matrices and correspond to $n$-flavors of fundamentals of
the gauge group $SU(\widetilde{N}_c-l)$ which is unbroken.
One can move $n$ D4-branes, from $(N_c'-l)$ flavor D4-branes stretched
between the $NS5_{-\theta}$-branes  and the $NS5_R'$-branes 
at $w=+v_{NS5_{-\theta}}
\cot \theta $, to the local minimum of the potential and the end
points of these $n$ D4-branes are at a nonzero $w$ \cite{GK0710-1}.
In the brane configuration from Figure 3, 
$\varphi$ and $\widetilde{\varphi}$ correspond to 
fundamental strings connecting between the $n$-flavor D4-branes and
$(\widetilde{N}_c-l)$-color D4-branes. Moreover,
the $h\Phi_n$ and ${ \varphi} 
\widetilde{\varphi}$
are $n \times n$ matrices.
The supersymmetric ground state corresponds to the vacuum expectation
values by
$h\Phi_n=\left[\frac{2\mu^2}{\mu_{\phi}(k+1)}\right]^{\frac{1}{k}}
{\bf 1}_{n}$ and $
\varphi \widetilde{\varphi} =0$. 

\begin{figure}[ht]
   \epsfxsize=2.5in 
\centerline{\epsffile{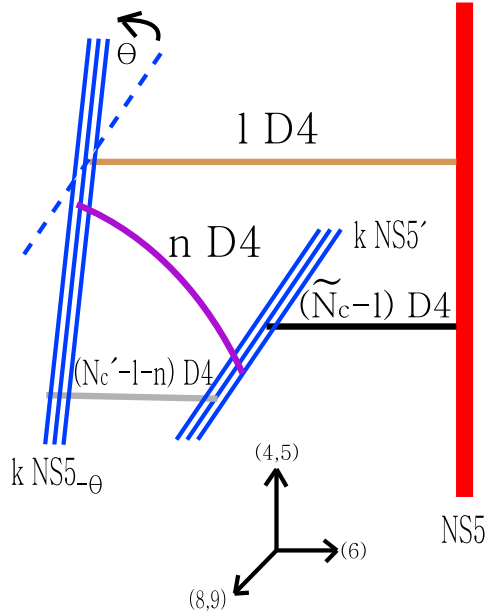}}
   \caption[FIG. \arabic{figure}.]{ 
The nonsupersymmetric meta-brane configuration 
corresponding to Figure 2 if we take into account  
the gravitational potential of NS5-brane. 
This can be done by moving $n$ flavor D4-branes 
from $(N_c'-l)$ flavor D4-branes of Figure 2 and 
the nonzero positive 
$w$ coordinate for $n$ curved flavor D4-branes can be
determined.
}
\end{figure}

The full one loop potential for 
$\Phi_n, \varphi$, and $\widetilde{\varphi}$
from (\ref{Dual}) and (\ref{Eigen}) including the one loop result
\cite{ISS} takes the form
\bea
\frac{V}{|h|^2}  =  
|\Phi_n  \varphi|^2   
+  |\Phi_n  \widetilde{\varphi}  |^2
  +  
\left| \varphi \widetilde{\varphi}-\mu^2 {\bf 1}_{n} + 
\frac{(k+1)h^k}{2} 
\mu_{\phi}  \Phi_n^k \right|^2 + b |h \mu|^2 \tr \Phi_n^{\dagger} \Phi_n, 
\label{pot}
\eea
where the positive numerical constant $b$ is given by
$b = \frac{(\ln 4-1)}{8\pi^2} (N_c'-N_c)$.
Differentiating this potential (\ref{pot}) with respect to 
$\Phi_n^{\dagger}$ and putting $\varphi=0=
\widetilde{\varphi}$, one obtains
\bea
\left[- \mu^2 {\bf 1}_{n}     +\frac{(k+1)h^k}{2} 
 \mu_{\phi} \Phi_n^k \right] \frac{k(k+1)}{2} (h^{\ast})^k
\mu_{\phi}^{\ast}
(\Phi_n^{\dagger})^{k-1}
 + b
|h \mu|^2 \Phi_n =0.
\label{equ}
\eea
Sine the higher order term superpotential plays the role of 
small perturbation and we assume that $\mu_{\phi}^{\frac{1}{2-k}} << \mu <<
\Lambda_m$ \cite{GK0710},
the second term of above will be negligible and 
one gets
\bea
  \frac{k(k+1)}{2} \mu^2 (h^{\ast})^k
\mu_{\phi}^{\ast}
(\Phi_n^{\dagger})^{k-1}
\simeq b
|h \mu|^2 \Phi_n.
\label{cond}
\eea
In order to see the vacuum structure, 
we consider the particular case where 
all the parameters are real including the $\Phi_n$ 
with $k >2$ \footnote{For $k=2$ case, one can express the expectation
value from (\ref{equ}) exactly: $h\Phi_n = \frac{2}{9} 
\frac{(3\mu_{\phi}-b) \mu^2 }{\mu_{\phi}^2}
{\bf 1}_n$.}.
The general case is, in principle, straightforward.
Then one arrives at the following form from (\ref{cond})
\bea
h \Phi_n 
\simeq \left[ \frac{2b }{k(k+1) \mu_{\phi}} \right]^{\frac{1}{k-2}}
{\bf 1}_n \qquad \mbox{or} \qquad
M_n \simeq \left[
\frac{\widetilde{N}_c}{\alpha k(k+1) 
\Lambda^3} \right]^{\frac{1}{k-2}} {\bf 1}_{n}.
\label{vac}
\eea
Then the vacuum energy $V$ is given by
$V \simeq n h^2 \mu^4$ and
expanding around this solution, one obtains
the eigenvalues for mass matrix for $\varphi$ and 
$\widetilde{\varphi}$: 
$
m_{\pm}^2
\simeq \left[\frac{2b}{k(k+1) 
\mu_{\phi}} \right]^{\frac{2}{k-2}}  \pm | h \mu|^2$.
For the positive value for these eigenvalues 
one should have
$
 \left[\frac{2b}{k(k+1) 
\mu_{\phi}} \right]^{\frac{1}{k-2}}  >   h \mu
$
which leads to 
the quarks $\varphi$ and $\widetilde{\varphi}$ are massive 
and then the vacuum (\ref{vac}) is locally stable
\footnote{The stability of the SUSY breaking vacua requires $
 \left[\frac{2b}{k(k+1) 
\mu_{\phi}} \right]^{\frac{1}{k-2}}  >   h \mu
$ and this impies that $h < 
\frac{\mu_{\phi}^{\frac{1}{2-k}}}{\mu} << 1$. As one flows to IR, the
dimensionless coupling $\mu_{\phi} E^{k-2}$(the mass dimension of
$\mu_{\phi}$ is equal to $2-k$) at the energy scale $E$
becomes smaller and smaller when $k > 2$. 
At the SUSY breaking scale $E=\sqrt{h}
\mu$, this dimensionless coupling becomes 
$h^{\frac{2-k}{2}} \mu_{\phi} (h \mu)^{k-2}$. When $k > 2$, the first
factor $h^{\frac{2-k}{2}}$ is large. Then the second factor 
should behave as 
$\mu_{\phi} (h \mu)^{k-2} << 1$ for the higher order deformation to
be  small. This is compatible with stability condition. However, 
the masses of $\varphi$ and $\widetilde{\varphi}$ become large when $k
> 2$
and are
integrated out. 
Let us emphasize here that the magnetic theory 
can stay the IR free region under the condition given in the footnote
2.
That is, $ \sqrt{h} \mu << \frac{1}{\mu_{\phi}^{k-2}}  <<  \left(
  \frac{\Lambda_1}{\mu}\right)^b
\Lambda_1 << \Lambda_1$. 
On the other hand, 
in the quartic case($ k=1$) where the quartic deformation becomes relevant
operator in the magnetic description, the qualitative difference  arises 
since the first factor $h^{\frac{2-k}{2}}
<< 1$, as long as $\frac{\mu_{\phi}}{h \mu}$ is not too large then 
the above requirement is valid. In
other words, the masses of $\varphi$ and $\widetilde{\varphi}$ remain
small enough compared to the SUSY breaking scale. 
}. 
Also note that $|h\Phi_n| << \Lambda_m$. 

It is evident that 
the $(N_c'-l-n)$ flavor D4-branes between 
the $NS5_{-\theta}$-branes and 
the $NS5_R'$-branes are related to the corresponding eigenvalues 
of $h\Phi$ (\ref{Eigen}), i.e.,  
$ 
\left[\frac{2\mu^2}{\mu_{\phi}(k+1)}\right]^{\frac{1}{k}} 
{\bf 1}_{N_c'-l-n}$ and 
the intersection point between the 
$(N_c'-l-n)$ D4-branes and the $NS5_R'$-branes is also given 
by $(v, w)=(0, +v_{NS5_{-\theta}} \cot \theta)$.
Moreover, 
the remnant $n$ flavor D4-branes between 
the $NS5_R'$-branes and the $NS5_{-\theta}$-branes 
are related to the corresponding eigenvalues 
(\ref{vac}) 
of $h\Phi_n$. 

$\bullet$ Separated $NS5_{-\theta}$-branes and $NS5_R'$-branes

Let us displace the $k$ $NS5_{-\theta}$-branes and $NS5_R'$-branes
given in Figure 2
in the $v$ direction respectively
to two $k$ different points denoted by 
$v_{NS5_{-\theta,j}}$ and $v_{NS5_{R,j}'}$
where $j=1,2, \cdots, k$.  
The color and flavor D4-branes attached to them are displaced as well.
The number of color D4-branes
stretched between $j$-th $NS5_{R,j}'$-brane and the NS5-brane
is denoted by $\widetilde{N}_{c,j}$
while the number of flavor D4-branes
stretched between the  $j$-th $NS5_{-\theta,j}$-brane and  the 
$j$-th $NS5_{R,j}'$-brane
is denoted by $N_{c,j}'$.
Many possible supersymmetric configurations, labeled by 
sets of nonnegative integers $(\widetilde{N}_{c,1},
\widetilde{N}_{c,2},
\cdots, \widetilde{N}_{c,k})$ and 
sets of nonnegative integers $(N_{c,1}',
N_{c,2}',
\cdots, N_{c,k}')$ with the initial conditions on the flavor- and
color- 
D4-branes where $l=0$
\bea
\sum_{j=1}^k \widetilde{N}_{c,j}(\equiv N_{c,j}'-N_{c,j})
= \widetilde{N}_c, \qquad
\sum_{j=1}^k N_{c,j}'= N_c'
\label{conforn}
\eea
can arise when there is no splitting of color and flavor D4-branes.
When all the $NS5_{R,j}'$-branes and 
$NS5_{-\theta,j}$-branes($j=1, 2, \cdots, k$) 
are distinct, the low energy  
physics corresponds to $k$ decoupled supersymmetric 
gauge theories with gauge groups 
\bea
\prod_{j=1}^{k} SU( \widetilde{N}_{c,j}) \times SU(N_{c,j}').
\nonu
\eea 
As we approach the origin of 
parameter space, $v_{NS5_{R,j}'}=0$ and
$v_{NS5_{-\theta,j}}=v_{NS5_{-\theta}}$ for all $j$, 
the full $SU(\widetilde{N}_c) \times SU(N_c')$
gauge group is restored.

Let us focus on $j$-th $NS5_{-\theta,j}$-brane, 
$j$-th $NS5_{R,j}'$-brane, $N_{c,j}'$ flavor D4-branes,
$\widetilde{N}_{c,j}$ color D4-branes and the NS5-brane.
Then the submeson field $M_j$ on this particular brane realization 
is defined as  $X_j \widetilde{X}_j$ and is 
in the representation  
$({\bf 1,{N_{c,j}^{\prime}}^2-1}) \oplus ({\bf 1, 1})$ under the 
gauge group $SU(\widetilde{N}_{c,j}) \times SU(N_{c,j}')$.
Note that
a bifundamental $X_j$ in an electric theory is in the representation 
$({\bf N_{c,j}, \overline{N_{c,j}'}})$ and its 
conjugate field $\widetilde{X}_j$ is  
in the representation $({\bf \overline{N_{c,j}}, N_{c,j}'})$. 
Generic $(v_{NS5_{-\theta,j}}-v_{NS5_{R,j}'})$, 
which is equal to $2\pi \ell_s^2 m_j$, 
is related to a polynomial superpotential for 
$M_j$. One can consider all the mass terms for the 
bifundamental for all $j$.
More explicitly, the expression  
$\left[m_j {\bf 1}_{N_{c,j}'} 
-\frac{\alpha(k+1)}{2} M_j^k\right]$(where $j=1,2, \cdots, k$) 
is nothing but the
derivative of superpotential with 
respect to the $M_j$ for $Y_j \widetilde{Y}_j=0$, from 
the F-term equations
(\ref{Fterm}), and has $k$ distinct minima 
$\left[\frac{2 m_j}{\alpha(k+1)}\right]^{\frac{1}{k}}$ plus zero
with some multiplicities. Note that the
F-term equations imply that there exist ``zero'' eigenvalues.
Here $Y_j$ and $\widetilde{Y}_j$ are
dual fields corresponding to $X_j$ and $\widetilde{X}_j$.

One can write 
the
derivative of superpotential, in terms of its eigenvalues $x$, with 
respect to the full meson $M$ for $Y \widetilde{Y}=0$ as
\bea
W_{dual}'(x) = \sum_{i=0}^{N_c'} s_i x^{N_c'-i} \equiv 
-\frac{\alpha (k+1)}{2} \prod_{j=1}^{k} x^{l_j} \left(x- 
\left[\frac{2 m_j}{\alpha(k+1)}\right]^{\frac{1}{k}} \right)^{N_{c,j}'-l_j}
\label{Wprime}
\eea
where we put the sum of multiplicities of zero eigenvalues as nonzero $l$ 
\bea
\sum_{j=1}^{k} l_j=l, \qquad 0 \leq l_j \leq \widetilde{N}_{c,j}.
\label{ljcon}
\eea 
The order of the polynomial (\ref{Wprime}) is $N_c'$.
The integers  $(N_{c,1}',
N_{c,2}',
\cdots, N_{c,k}')$ are the number of the nonzero plus zero 
eigenvalues of the meson
$M$ residing in the different minima.
Thus the set of  
\bea
(N_{c,1}',
N_{c,2}',
\cdots, N_{c,k}'), \qquad (l_1, \l_2, \cdots,
l_k), \qquad \mbox{and} \qquad
(m_1, m_2, \cdots, m_k)
\nonu
\eea
determines the expectation value of $M$ completely. 
In the brane realization of Figure 4(for the time being, the further
splitting of $n_j$ flavor D4-branes is ignored), 
the first two characterize how to split the
flavor D4-branes and the last shows the relative distance between 
the $NS5_{-\theta,j}$-brane and the $NS5_{R,j}'$-brane along $v$ direction. 

\begin{figure}[ht]
   \epsfxsize=2.5in 
\centerline{\epsffile{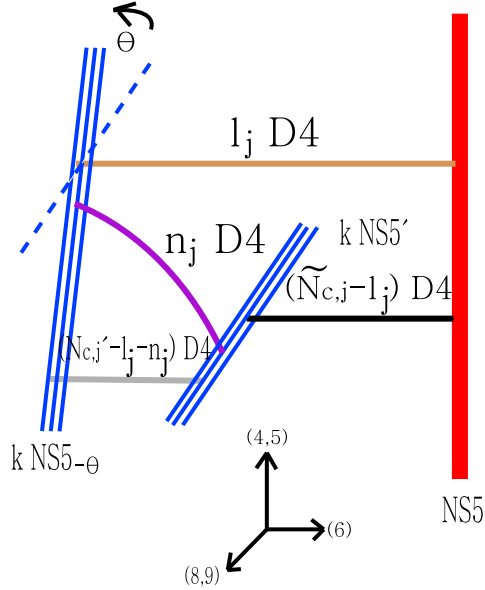}}
   \caption[FIG. \arabic{figure}.]{ 
The nonsupersymmetric meta-brane configuration 
corresponding to Figure 2 when the $NS5_{-\theta}$-branes 
and the $NS5_{R}'$-branes are separated. 
We only draw $(N_{c,j}'-l_j-n_j)$-flavor D4-branes, $n_j$-flavor
D4-branes, 
connecting between 
$j$-th $NS5_{-\theta,j}$-brane and 
$j$-th $NS5_{R,j}'$-brane,
$l_j$ D4-branes connecting between 
 $j$-th $NS5_{-\theta,j}$-brane and
the NS5-brane,
and 
$(\widetilde{N}_{c,j}-l_j)$-color D4-branes connecting between  
$j$-th $NS5_{R,j}'$-brane and the NS5-brane. Other $(k-1)$ possible
brane configurations we do not draw are assumed.}
\end{figure}

Since the eigenvalues for the submeson field $M_j$
are either $0$ or
$\left[\frac{2m_j}{\alpha(k+1)}\right]^{\frac{1}{k}}$, 
one takes
$N_{c,j}' \times N_{c,j}'$ matrix $M_j$ with $l_j$'s 
eigenvalues $0$ and $(N_{c,j}'-l_j)$'s
eigenvalues $\left[\frac{2m_j}{\alpha(k+1)}\right]^{\frac{1}{k}}$ as follows:
\bea
M = \left[\frac{2}{\alpha(k+1)}\right]^{\frac{1}{k}} \left(
\begin{array}{ccccc}
0{\bf 1}_l & 0 & 0 & \cdots  & 0 \\
0 & m_1^{\frac{1}{k}} 
{\bf 1}_{N_{c,1}'-l_1} & 0 & \cdots & 0  \\
\cdot & \cdot & \cdot & \cdots & 0 \\
0 & 0 &  0 & 0 & m_k^{\frac{1}{k}} 
{\bf 1}_{N_{c,k}'-l_{k}}  
\end{array}
\right) = \left(
\begin{array}{ccc}
M_1 & 0 & 0  \\
0 & M_2  & 0  \\
\cdot & \cdot & \cdot \\
0 & 0 & M_k
\end{array}
\right)
\label{M01}
\eea
with (\ref{ljcon}) and (\ref{conforn}).
In the last equation (\ref{M01}) 
we redistributed the zeros of $0{\bf 1}_l$ along the diagonal
direction into
each 
submeson $M_j$ appropriately. This structure is evident from the brane 
configuration.
In the brane configuration of Figure 4, the $l_j$ of the
$N_{c,j}'$-flavor D4-branes are connected with $l_j$ of $\widetilde{N}_{c,j}$-color
D4-branes
and the resulting $l_j$ D4-branes 
stretch from the $NS5_{-\theta,j}$-brane to
the NS5-brane directly 
and the intersection point between the 
$l_j$ D4-branes and the NS5-brane is $(v, w)=(+v_{NS5_{-\theta,j}}, 0)$.
This corresponds to  exactly the $l_j$'s eigenvalues $0$ of 
$M_j$ in (\ref{M01}).
Now the remaining $(N_{c,j}'-l_j)$-flavor D4-branes between 
the $NS5_{-\theta,j}$-brane and 
the $NS5_{R,j}'$-brane correspond to the remaining eigenvalues 
of $M_j$ in (\ref{M01}), i.e.,   
$\left[\frac{2m_j}{\alpha(k+1)}\right]^{\frac{1}{k}} {\bf 1}_{N_{c,j}'-l_j}$.
The intersection point between the 
$(N_{c,j}'-l_j)$ D4-branes and the $NS5_{R,j}'$-brane is given 
by $(v, w)=(0, +v_{NS5_{-\theta,j}} \cot \theta)$ from the geometry.

By substituting the above $M$ (\ref{M01}) into the F-term equation, 
one also obtains
\bea
Y  \widetilde{Y} = \left(
\begin{array}{cccccc}
-m_1 \Lambda_1 {\bf 1}_{l_1} & 0 & 0 & \cdots & 0  & 0  \\
\cdot  & \cdot & \cdot & \cdots & \cdot  & 0 \\
0 & 0 & 0 & \cdots & -m_k \Lambda_k {\bf 1}_{l_k}  & 0 \\ 
0 & 0 & 0 & 0 & 0 & 0 {\bf 1}_{N_c'-l}  
\end{array}
\right)
 = \left(
\begin{array}{ccc}
Y_1 \widetilde{Y}_1 & 0 & 0  \\
0 & Y_2 \widetilde{Y}_2  & 0  \\
\cdot & \cdot & \cdot \\
0 & 0 & Y_k \widetilde{Y}_k
\end{array}
\right).
\label{solqq11}
\eea
In the last equation, 
we redistributed the zeros of $0{\bf 1}_{N_c'-l}$ into
each $Y_j \widetilde{Y}_j$.
In the $l_j$-th vacuum the gauge symmetry is broken to $SU(\widetilde{N}_{c,j}-l_j)$
and 
the supersymmetric vacuum drawn in Figure 4 with $l_j=0$ has 
$Y_j  \widetilde{Y}_j =0$
and the gauge group 
$SU(\widetilde{N}_{c,j})$ is unbroken. 

So far, the ground states are supersymmetric. Moreover,
the theory has many nonsupersymmetric meta-stable ground states.
One can deform the Figure 3 by displacing the multiple NS-branes 
along $v$ direction.
Then the $n$ curved flavor D4-branes attached to them as well as other
D4-branes are displaced
also as $k$ different $n_j$'s connecting between $NS5_{-\theta,j}$-brane and
$NS5_{R,j}'$-brane.
For the IR free region, $N_{c,j} < N_{c,j}' < \frac{3}{2} N_{c,j}$ \cite{ISS}, 
the magnetic theory is the effective low energy description of the
asymptotically free electric gauge theory.
When we rescale the submeson field as
$M_{j} = h \Lambda \Phi_{j} $,
then the Kahler potential for $\Phi_{j}$ is canonical and the magnetic
quarks $Y_j$ and $\widetilde{Y}_j$ 
are canonical near the origin of field space.
Then the magnetic superpotential (\ref{superpo11}) 
can be rewritten in terms of $\Phi_j$ and $Y_j \widetilde{Y}_j$
\bea
W_{dual} = \sum_{j=1}^{k} \left[
 h \Phi_j  Y_j   \widetilde{Y}_j 
 +  
\frac{ \mu_{\phi}}{2} h^{k+1} \tr 
\Phi_j^{k+1}- h \tr \mu_j^2  \Phi_j \right]
\label{dDual}
\eea
with
$
\mu_j^2 = -m_j \Lambda_j$ and 
$\mu_{\phi} = -\alpha \Lambda^{k+1}$ as before. 
Remember that both $ h \Phi_j$ and $ Y_j   \widetilde{Y}_j$ are 
$N_{c,j}' \times N_{c,j}'$ matrices and the $\widetilde{N}_{c,j}$
indices are contracted in the product  $ Y_j   \widetilde{Y}_j$ in the
cubic term of (\ref{dDual}).
One splits 
the $(N_{c,j}'-l_j) \times (N_{c,j}'-l_j)$
block  at the lower right corner of $h\Phi_j$ (\ref{M01}) and $Y_j
\widetilde{Y}_j$ (\ref{solqq11}) 
into blocks of 
size $n_j$ and $(N_{c,j}'-l_j-n_j)$ for all $j$ as follows \cite{GK0710}:
\bea
h \Phi & = & \left[\frac{2}{\mu_{\phi}(k+1)}\right]^{\frac{1}{k}} \left(
\begin{array}{ccccc}
0 {\bf 1}_{l+n} & 0 & 0 & \cdots  & 0 \\
0 & \mu_1^{\frac{2}{k}} 
{\bf 1}_{N_{c,1}'-l_1-n_1} & 0 & \cdots & 0  \\
\cdot & \cdot & \cdot & \cdots & 0 \\
0 & 0 &  0 & 0 & \mu_k^{\frac{2}{k}} 
{\bf 1}_{N_{c,k}'-l_{k}-n_k}  
\end{array}
\right) \nonu \\
& + & 
\mbox{diag} (0 {\bf 1}_{l}, h \Phi_{n_1}, \cdots, h \Phi_{n_k},
0 {\bf 1}_{N_c'-l-n})
\label{M02}
\eea
and
\bea
Y  \widetilde{Y}   & = &  \left(
\begin{array}{cccccc}
\mu_1^2  {\bf 1}_{l_1} & 0 & 0 & \cdots & 0  & 0  \\
\cdot  & \cdot & \cdot & \cdots & \cdot  & 0 \\
0 & 0 & 0 & \cdots & \mu_k^2  {\bf 1}_{l_k}  & 0 \\ 
0 & 0 & 0 & 0 & 0 & 0 {\bf 1}_{N_c'-l} 
\end{array}
\right) \nonu \\
& + &   \mbox{diag} (
0{\bf 1}_{l}, \varphi_{n_1} \widetilde{\varphi}_{n_1}, 
\cdots,  \varphi_{n_k} \widetilde{\varphi}_{n_k}, 
0 {\bf 1}_{N_c'-l-n}).
\label{solqq2}
\eea
Here 
$\varphi_{n_j}$ and $\widetilde{\varphi}_{n_j}$ are 
$n_j \times (\widetilde{N}_{c,j}-l_j)$
matrices and correspond to $n_j$-flavors of fundamentals of
the gauge group $SU(\widetilde{N}_{c,j}-l_j)$ which is unbroken.
Let us denote the sum of $n_j$ by
\bea
\sum_{j=1}^k n_j = n, \qquad 0 \leq n_j \leq N_{c,j}'-l_j
\nonu
\eea
which is the number of curved D4-branes in Figure 3 before deformation
we are considering. 
In the brane configuration from Figure 4, 
they correspond to 
fundamental strings connecting between the $n_j$-flavor D4-branes and
$(\widetilde{N}_{c,j}-l_j)$-color D4-branes. Moreover,
the $\Phi_{n_j}$ and ${ \varphi}_{n_j} 
\widetilde{\varphi}_{n_j}$
are $n_j \times n_j$ matrices.
The supersymmetric ground state corresponds to the vacuum expectation
values by
$h\Phi_{n_j}=\left[\frac{2\mu_j^2}{\mu_{\phi}(k+1)}\right]^{\frac{1}{k}}
{\bf 1}_{n_j}$ and $
\varphi_{n_j} \widetilde{\varphi}_{n_j}=0$.
Note that the $\Phi_j$ contains $\Phi_{n_j}$ as a submatrix.

The full one loop potential for 
$\Phi_{n_j}, \varphi_{n_j}$, and $\widetilde{\varphi}_{n_j}$
from (\ref{dDual}), (\ref{M02}) and (\ref{solqq2}) takes the form
\bea
\frac{V}{|h|^2}  =  \sum_{j=1}^k \left[
|\Phi_{n_j}  \varphi_{n_j}|^2   
+  |\Phi_{n_j}  \widetilde{\varphi}_{n_j}  |^2
  +  
\left| \varphi_{n_j} \widetilde{\varphi}_{n_j}-\mu_j^2 {\bf 1}_{n_{j}} + 
\frac{(k+1)h^k}{2} 
\mu_{\phi}  \Phi_{n_j}^k \right|^2 + b_j |h \mu_j|^2 \tr \Phi_{n_j}^{\dagger}
\Phi_{n_j} 
\right], 
\nonu
\eea
where
$b_j = \frac{(\ln 4-1)}{8\pi^2} (N_{c,j}'-N_{c,j})$ for all $j$ 
\cite{ISS}. The first term 
is the absolute valued square of derivative of the superpotential 
(\ref{dDual}) with respect to the field $\widetilde{\varphi}_{n_j}$,
the second term to the field $\varphi_{n_j}$, and 
the third term to the field $\Phi_{n_j}$. 
The last term is due to the one loop potential. 
Differentiating this potential with respect to 
$\Phi_{n_j}^{\dagger}$ and putting $\varphi_{n_j}=0=
\widetilde{\varphi}_{n_j}$, one obtains
\bea
\left[- \mu_j^2 {\bf 1}_{n_j}     +\frac{(k+1)h^k}{2} 
 \mu_{\phi} \Phi_{n_j}^k \right] \frac{k(k+1)}{2} (h^{\ast})^k
\mu_{\phi}^{\ast}
(\Phi_{n_j}^{\dagger})^{k-1}
 + b_j
|h \mu_j|^2 \Phi_{n_j} =0.
\nonu
\eea
Sine the higher order term superpotential plays the role of 
small perturbation and we assume that $\mu_{\phi}^{\frac{1}{k-2}} << \mu_j <<
\Lambda_{m,j}$ \cite{GK0710},
the second term of above will be negligible and 
one gets
\bea
  \frac{k(k+1)}{2} \mu_j^2 (h^{\ast})^k
\mu_{\phi}^{\ast}
(\Phi_{n_j}^{\dagger})^{k-1}
\simeq b_j
|h \mu_j|^2 \Phi_{n_j}.
\nonu
\eea
In order to see the structure of this solution, 
we consider the particular case where 
all the parameters are real including the $\Phi_{n_j}$ with $k >2$.
The general case is, in principle, straightforward.
Then one arrives at the following form
\bea
h \Phi_{n_j} 
\simeq \left[ \frac{2b_j }{k(k+1) \mu_{\phi}} \right]^{\frac{1}{k-2}}
{\bf 1}_{n_j} \qquad \mbox{or} \qquad
M_{n_j} \simeq \left[
\frac{\widetilde{N}_{c,j}}{\alpha k(k+1) 
\Lambda_j^3} \right]^{\frac{1}{k-2}} {\bf 1}_{n_j}.
\label{vac1}
\eea
Of course, the expression 
$\sum_{j=1}^k \tr (h \Phi_{n_j})^{k-2}$, when $b_j =\frac{b}{k}$ and
$n_j=\frac{n}{k}$, 
reduces to $\frac{1}{k} \tr (h \Phi_{n})^{k-2}$ 
with (\ref{vac}).
Then the vacuum energy $V$ is given by
$V \simeq \sum_{j=1}^k n_j h^2 \mu_j^4$ and
expanding around this solution, one obtains
the eigenvalues for mass matrix for $\varphi_{n_j}$ and 
$\widetilde{\varphi}_{n_j}$: 
$
m_{\pm}^2
\simeq \left[\frac{2b_j}{k(k+1) 
\mu_{\phi}} \right]^{\frac{2}{k-2}}  \pm | h \mu_j|^2$.
For the positive value for these, 
one should have
$
 \left[\frac{2b_j}{k(k+1) 
\mu_{\phi}} \right]^{\frac{1}{k-2}}  >   h \mu_j
$
which leads to 
the quarks $\varphi_{n_j}$ and $\widetilde{\varphi}_{n_j}$ are massive 
and then the vacuum (\ref{vac1}) is locally stable. 

The $(N_{c,j}'-l_j-n_j)$ flavor D4-branes between 
the $NS5_{-\theta,j}$-brane and 
the $NS5_{R,j}'$-brane are related to the corresponding eigenvalues 
of $h\Phi$ (\ref{M02}), i.e.,  
$ 
\left[\frac{2\mu_j^2}{\mu_{\phi}(k+1)}\right]^{\frac{1}{k}} 
{\bf 1}_{N_{c,j}'-l_j-n_j}$ and 
the intersection point between the 
$(N_{c,j}'-l_j-n_j)$ D4-branes and the $NS5_{R,j}'$-brane is also given 
by $(v, w)=(0, +v_{NS5_{-\theta,j}} \cot \theta)$.
Moreover, 
the remnant $n_j$ flavor D4-branes between 
the $NS5_{R,j}'$-brane and the $NS5_{-\theta,j}$-brane 
are related to the corresponding eigenvalues 
(\ref{vac1}) 
of $h\Phi_{n_j}$ and corresponding nonzero $w$ coordinate of these
curved flavor D4-branes can be determined. 

Therefore, the meta-stable states, for fixed $k$ which is related to 
the order of the superpotential polynomial and $\theta$ which is a
deformation parameter by rotation angle of $NS5_{-\theta,j}$-brane, 
are classified by the number of various D4-branes and the positions of
multiple NS-branes: 
\bea
(N_{c,j}, N_{c,j}', l_j, n_j) \qquad \mbox{and} \qquad
(v_{NS5_{-\theta,j}}, v_{NS5_{R,j}'}).
\nonu
\eea
That is, the former is given by 
$i)$ $N_{c,j}$ which appears in the number of dual color D4-branes
through $\widetilde{N}_{c,j}$ and  in $h \Phi_{n_j}$ (\ref{vac1}) through
$b_j$,  
$ii)$ $N_{c,j}'$ which is encoded also in 
the number of dual color D4-branes through $\widetilde{N}_{c,j}$, 
in ``straight'' flavor D4-branes, in the
multiplicities of expectation value $h\Phi$ (\ref{M02}), and  in
$h \Phi_{n_j}$ through $b_j$, 
$iii)$ $l_j$
which characterizes the number of splitting D4-branes
and the
multiplicities of expectation value $h\Phi$, 
and $iv)$ $n_j$ which is
the number of ``curved'' flavor D4-branes and determines the
multiplicities of the expectation value 
for $h \Phi_{n_j}$  and the
multiplicities of expectation value $h\Phi$. 
The latter is given by 
$i)$ $v_{NS5_{-\theta,j}}$ which provides the locations of both $l_j$ D4-branes
and stright flavor D4-branes through $\mu_j$ (\ref{M02}), and 
$ii)$ $v_{NS5_{R,j}'}$ which determines the locations of both dual 
color D4-branes and  straight flavor  
D4-branes through $\mu_j$. 

\section{Meta-stable brane configuration with $(4k+3)$ NS-branes plus O4-plane }

\subsection{Electric theory}

The type IIA brane configuration \cite{Ahn07-5} corresponding to 
${\cal N}=1$ supersymmetric gauge theory with
gauge group
$
Sp(N_c) \times SO(2N_c')
$
and a bifundamental $X$ in the representation 
$({\bf 2N_c, 2N_c'})$   
can be described by 
a middle NS5-brane,
the left $(2k+1)$ $NS5_L'$-branes and the right 
$(2k+1)$ $NS5_R'$-branes, $2N_c$-  and $2N_c'$-color D4-branes, 
and an
O4-plane(01236). 
The O4-plane acts as $(x^4,x^5,x^7,x^8,x^9) \rightarrow
(-x^4,-x^5,-x^7,
-x^8,-x^9)$ as usual \cite{Tatar,Ahn97}.
The number of $NS5_{L,R}'$-branes 
is a nonnegative integer $k \geq 0$.
The $2N_c$ D4-branes and $O4^{+}$-plane 
are suspended between the middle 
NS5-brane whose $x^6$ coordinate is $x^6=0$ 
and $NS5_R'$-branes while 
the $2N_c'$ D4-branes and $O4^{-}$-plane 
are suspended between the $NS5_L'$-branes and the middle NS5-brane.
We take the arbitrary numbers of color D4-branes with the constraint 
$N_c' \geq N_c+2$.

Let us deform this gauge theory. 
Displacing the two kinds of $NS5_{L,R}'$-branes relative each other in
the $\pm v$ 
direction corresponds to turning on a quadratic
mass-deformed superpotential
for the bifundamental $X$  
while rotating them in the $(v,w)$ plane corresponds to 
turning on a quartic($k=0$) or higher order($k \geq 1$)  
superpotential for the bifundamental $X$.
The deformed electric superpotential, by changing the left $(2k+1)$
$NS5_{L}'$-branes, 
is 
as follows:
\bea
W_{elec} = -\frac{\alpha}{2}   
\tr (X X)^{2k+2} +  \tr m X X,
\qquad
\alpha = \frac{\tan \theta}{\Lambda}, 
\qquad m = \frac{v_{NS5_{-\theta}}}{2\pi \ell_s^2}. 
\label{superelectric1}
\eea
Each $(k+\frac{1}{2})$ $NS5_L'$-branes are moving to the $ \pm v$
directions respectively together with $N_c'$
D4-branes and are rotating 
by an angle $-\theta$ in $(w,v)$-plane.
Then the 
$v$ coordinate of $NS5_L'$-branes is denoted by $v= \pm
v_{NS5_{-\theta}}$. The order of this superpotential
(\ref{superelectric1}) 
can be obtained by replacing $k$ in (\ref{superelectric}) with
$(2k+1)$. 
We present the electric brane configuration in Figure 5.

\begin{figure}[ht]
   \epsfxsize=2.5in 
\centerline{\epsffile{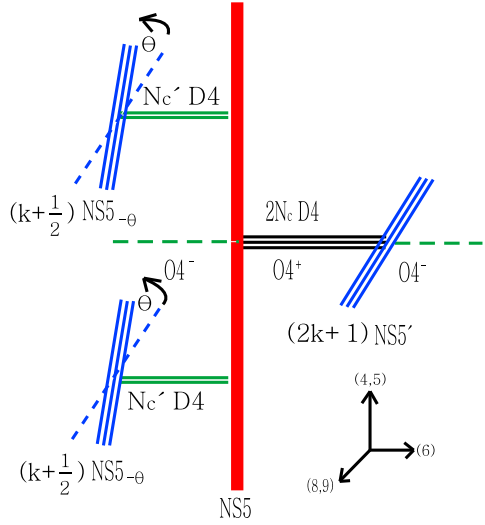}}
   \caption[FIG. \arabic{figure}.]{ 
The  ${\cal N}=1$ supersymmetric 
electric brane configuration for the gauge group $Sp(N_c) \times
SO(2N_c')$ and a bifundamental $X$  with nonvanishing mass and higher 
order terms 
for the bifundamental. 
There are two deformations by  
rotation  and displacement of upper and lower $(k+\frac{1}{2})$ 
$NS5_L'$-branes. Note that there exist multiple outer NS-branes.}
\end{figure}

The solution for the supersymmetric vacua can be written as 
$X X =\left[\frac{m}{\alpha(k+1)}\right]^{\frac{1}{2k+1}}$ 
through the F-term condition.
This breaks 
the gauge group $Sp(N_c) \times SO(2N_c')$ 
to $Sp(p_0)$, $SO(2N_c'-2N_c+2p_0)$,
and $U(2p_1) \times U(2p_2) \times \cdots \times U(2p_k)$ with
2$\sum_{i=0}^k p_i =2N_c$ \cite{ILS}.
The $2p_0$ is the number of eigenvalues which are zero with 
$X=0$.
When the middle NS5-brane moves to $ \pm w$ direction, then the three
kinds of NS-branes intersect in three points 
in $(v,w)$-plane(and their mirrors). 
Then we consider that $2p_0$
D4-branes are connecting between the middle NS5-brane and the zero-th
$NS5_{R,0}'$-brane. One can decompose $2N_c$ D4-branes as 
$2p_0$ plus $2(N_c-p_0)$ D4-branes.
The latter can be reconnected with the same number of D4-branes 
among $2N_c'$ D4-branes.
Then the remaining $2(N_c'-N_c+p_0)$
D4-branes are connecting between the zero-th $NS5_{-\theta, 0}$-brane and 
the middle NS5-brane. 
Then $2(N_c-p_0)$ D4-branes are suspended between $2k$
$NS5_{-\theta}$-branes and $NS5_R'$-branes.
By separating these NS-branes transversely, each $p_i$
D4-branes(where $i=1, 2, \cdots, k$) 
are connecting between the $i$-th $NS5_{-\theta, i}$-brane and the
$i$-th $NS5_{R,i}'$-brane directly(and their mirrors).

\subsection{Magnetic theory}

In order to apply the Seiberg dual to the $Sp(N_c)$ factor,  
the NS5-brane is moved to the right all the
way past the $NS5_R'$-branes. Let us
introduce $2N_c'$ D4-branes and $2N_c'$ 
anti-D4-branes  between $NS5_R'$-branes and   
NS5-brane, recombine the former with  
the $2N_c'$ D4-branes that are connecting between 
the $NS5_{-\theta}$-branes and the $NS5_R'$-branes. By 
moving those combined D4-branes
to $\pm v$-direction, 
one gets the final Figure 6(with $l=0$).

\begin{figure}[ht]
   \epsfxsize=2.5in 
\centerline{\epsffile{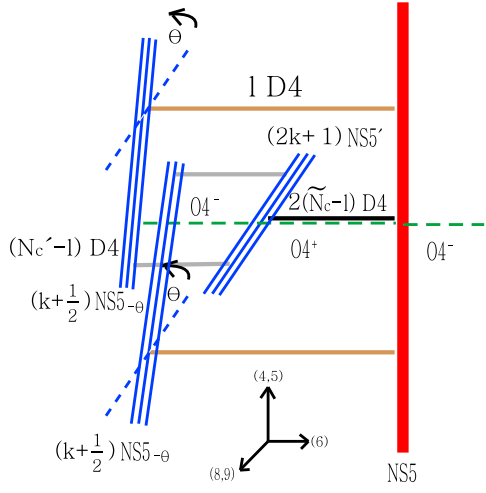}}
   \caption[FIG. \arabic{figure}.]{ 
The  ${\cal N}=1$ supersymmetric 
magnetic brane configuration corresponding to Figure 5 with 
a misalignment between D4-branes if we ignore the gravitational
potential of NS5-brane. 
The upper $N_c'$-flavor D4-branes connecting between
the upper half 
$NS5_{-\theta}$-brane and $NS5_R'$-brane are splitting into $(N_c'-l)$ and
$l$ D4-branes. 
The location of intersection between the upper  $NS5_{-\theta}$-branes
and the upper $(N_c'-l)$
D4-branes 
is given by $(v,w)=(0, +v_{NS5_{-\theta}} 
\cot \theta)$ while the one between  
the upper $NS5_{-\theta}$-brane and the upper $l$
D4-branes 
is given by $(v,w)=(+v_{NS5_{-\theta}}, 0)$.}
\end{figure}

The gauge group is given by
$
Sp(\widetilde{N}_c=N_c'-N_c-2) \times SO(2N_c')
$
and 
the matter contents are the field $Y$ in the 
representation $({\bf 2\widetilde{N}_c, 2N_c'})$ under the dual gauge
group 
and  the gauge-singlet $M(\equiv X X)$ is
in the representation
$({\bf 1, N_c'(2N_c'-1)})$ under 
the dual gauge group.
Then the dual magnetic superpotential
\footnote{
The conditions
 $b_{Sp(\widetilde{N}_c)}^{mag}< 0$ and 
$b_{Sp(N_c)} > 0$ imply
that $N_c' < \frac{3}{2} N_c +\frac{3}{2}$. 
Then the range for the $N_c'$ 
can be written as $ N_c +2 < N_c' 
< \frac{3}{2} N_c +\frac{3}{2} $. 
Moreover, $b_{SO(2N_c')} > 0$ and 
$b_{SO(2N_c')}^{mag}> 0$.
In the low energy description, the $SO(2N_c')$ is asymptotically free.
 At the scale Landau pole $\Lambda_1$, the $Sp(N_c)$ theory is strongly
coupled and the Seiberg duality occurs. The coefficients of beta function 
$b_{Sp(\widetilde{N}_c)}^{mag}$ becomes negative and 
$b_{SO(2N_c')}^{mag}$
becomes positive. 
Then at energy scale lower than $\Lambda_1$, 
the theory is weakly coupled. 
We require that 
$SO(2N_c')^{mag}$ theory is less coupled than the 
$Sp(\widetilde{N}_c)^{mag}$ at the supersymmetry breaking scale $\mu$.
This 
provides a stronger constraint on $\Lambda_2$. 
Then under the constraint, $\Lambda_2 <<  \left(
  \frac{\Lambda_1}{\mu}\right)^b
\Lambda_1 << \Lambda_1$
where $b \equiv 
\frac{b_{Sp(\widetilde{N}_c)}^{mag}-b_{SO(2N_c')}^{mag}}{b_{SO(2N_c')}}$, 
one can ignore the contribution from the gauge coupling of 
$SO(2N_c')^{mag}$ at the supersymmetry breaking scale and one relies on
the one loop computation.}, 
by adding the mass term
and higher order term for the bifundamental $X$ (\ref{superelectric1}) 
to the cubic
superpotential, is given by
\bea
W_{dual} =  \frac{1}{\Lambda} \tr M Y Y- 
\frac{\alpha}{2} \tr M^{2k+2}  + \tr m M
\label{supernew}
\eea
and the F-term equations are
\bea
 M Y  =  0, \qquad 
-\frac{1}{\Lambda} Y Y & = & m -\alpha(k+1) M^{2k+1}. 
\label{Fterm1}
\eea

Let us consider all the 
$NS5_{-\theta}$-branes and $NS5_R'$-branes
are coincident with each other. 

$\bullet$ Coincident $NS5_{-\theta}$-branes and $NS5_R'$-branes

The matrix equation  
\footnote{The mass matrix $m$
is antisymmetric in the indices and is given by $m =
\mbox{diag}(i\sigma_2 m_1, i\sigma_2 m_2, \cdots, i\sigma_2
m_{N_c'})$ due to the antisymmetric matrix $M$. In the matrix equation
$m M$, we assumed this property of mass matrix. In (\ref{M0new}), we
use the same notation for the equal mass 
$m \equiv m_1=m_2=\cdots =m_{N_c'}$.}
$
m M = \alpha (k+1) M^{2(k+1)}$ implies
that the eigenvalues for the meson field $M$ 
are either $0$ or
$\left[\frac{m}{\alpha(k+1)}\right]^{\frac{1}{2k+1}}$, and 
one takes
$2N_c' \times 2N_c'$ matrix with $2l$'s eigenvalues $0$ and $2(N_c'-l)$'s
eigenvalues $\left[\frac{m}{\alpha(k+1)}\right]^{\frac{1}{2k+1}}$:
\bea
M = \left(
\begin{array}{cc}
0 {\bf 1}_{2l} & 0  \\
0 & \left[\frac{m}{\alpha(k+1)}\right]^{\frac{1}{2k+1}} {\bf 1}_{N_c'-l}
\otimes i \sigma_2 
\end{array}
\right)
\label{M0new}
\eea
where $l=1, 2, \cdots, 2N_c'$. 
Therefore, in the brane configuration of Figure 6, the $l$ of the
upper $N_c'$ flavor D4-branes are connected with $l$ of $\widetilde{N}_c$ color
D4-branes
and the resulting D4-branes stretch from the upper $NS5_{-\theta}$-branes to
the NS5-brane directly and the intersection point between the 
$l$ upper D4-branes and the NS5-brane is given by $(v,
w)=(+v_{NS5_{-\theta}}, 0)$.
Similarly the mirrors are located at  $(v,
w)=(-v_{NS5_{-\theta}}, 0)$.
This corresponds to  exactly the $2l$'s eigenvalues 
$0$ of $M$ above (\ref{M0new}).
The remaining $(N_c'-l)$ upper flavor D4-branes between 
the $NS5_{-\theta}$-branes and 
the $NS5_R'$-brane are related to the corresponding half eigenvalues 
of $M$ which is equal to  
$ \left[\frac{m}{\alpha(k+1)}\right]^{\frac{1}{2k+1}} 
{\bf 1}_{N_c'-l} \otimes i \sigma_2$.
The intersection point between the 
$(N_c'-l)$ upper D4-branes and the $NS5_R'$-branes is given 
by $(v, w)=(0, +v_{NS5_{-\theta}} \cot \theta)$ corresponding to
half eigenvalues of $M$ from geometry.
The mirrors are located at  
$(v, w)=(0, -v_{NS5_{-\theta}} \cot \theta)$ corresponding to other half
eigenvalues of $M$ \cite{Ahn08-1two}.

Substituting (\ref{M0new}) into the second equation of (\ref{Fterm1})
gives rise to 
\bea
Y Y = \left(
\begin{array}{cc}
-m \Lambda {\bf 1}_{2l} & 0  \\
0 & 0 {\bf 1}_{2N_c'-2l}
\end{array}
\right).
\label{solqq1}
\eea
In the $l$-th vacuum the gauge symmetry is broken to $Sp(\widetilde{N}_c-l)$
and 
the supersymmetric vacuum drawn in Figure 6 with $l=0$ has 
$Y=0$ and the gauge group 
$Sp(\widetilde{N}_c)$ is unbroken. The expectation value of $M$
(\ref{M0new}) 
in this case 
is given by
$M = \left[\frac{m}{\alpha(k+1)}\right]^{\frac{1}{2k+1}}  
{\bf 1}_{N_c'} \otimes i \sigma_2
= \left[ \frac{m \Lambda \cot \theta}{(k+1)} \right]^{\frac{1}{2k+1}} 
{\bf 1}_{N_c'} \otimes i \sigma_2$.
If we replace $(2k+1)$ $NS5_{-\theta}$-branes with $2N_f$ D6-branes
and  $(2k+1)$ $NS5_{R}'$-branes with a single $NS5_R'$-brane, then 
the meta-stable brane configuration of Figure 6 reduces to the one in 
\cite{Ahn06-1,Ahn07-2,Ahn07-8}.

The theory has many nonsupersymmetric meta-stable ground 
states if  an attractive
gravitational interaction
between the flavor D4-branes and the NS5-brane from 
the DBI action is considered \cite{GK}.
For the IR free region \cite{ISS,Ahn06-1}, 
the magnetic theory is the effective low energy description of the
asymptotically free electric gauge theory.
When we rescale the meson field as
$M = h \Lambda \Phi $,
then the Kahler potential for $\Phi$ is canonical and the magnetic
``quarks'' are canonical near the origin of field space.
Then the magnetic superpotential (\ref{supernew}) 
can be written in terms of $\Phi$
\bea
W_{dual} = 
 h \Phi  Y   Y 
 +  
\frac{ \mu_{\phi}}{2} h^{2k+2} \tr \Phi^{2k+2}- h \mu^2 \tr \Phi 
\label{Dualnew}
\eea
with
$
\mu^2 = -m \Lambda$ and $ 
\mu_{\phi} = -\alpha \Lambda^{2k+2}$.

The classical supersymmetric vacua given by (\ref{M0new}) and
(\ref{solqq1})
can be described in terms of new variables.
Now one splits 
the $2(N_c'-l) \times 2(N_c'-l)$
block  at the lower right corner of $h\Phi$ and $Y Y$ 
into blocks of 
size $2n$ and $2(N_c'-l-n)$ as follows:
\bea
h\Phi = \left(
\begin{array}{ccc}
0{\bf 1}_{2l} & 0 & 0  \\
0 & h \Phi_{2n} & 0 \\
0 & 0 & \left[\frac{\mu^2}{\mu_{\phi}(k+1)}\right]^{\frac{1}{2k+1}} 
{\bf 1}_{N_c'-l-n} \otimes i \sigma_2
\end{array}
\right), 
Y^2 = \left(
\begin{array}{ccc}
\mu^2 {\bf 1}_{2l} & 0 & 0  \\
0 & \varphi \varphi  &  0 \\
0 & 0 & 0{\bf 1}_{2N_c'-2l-2n}
\end{array}
\right)
\label{Eigennew}
\eea
where $\varphi$  is $2n \times 2(\widetilde{N}_c-l)$
matrix and corresponds to $2n$ flavors of fundamentals of
the gauge group $Sp(\widetilde{N}_c-l)$ which is unbroken by the nonzero
expectation value of $Y$.
One can move $n$ upper D4-branes, from upper $(N_c'-l)$ D4-branes stretched
between the $NS5_R'$-brane and the upper $NS5_{-\theta}$-brane 
at $w=+v_{NS5_{-\theta}}
\cot \theta $, to the local minimum of the potential and the end
points of these $n$ D4-branes are at a nonzero $w$ as in Figure 7.
In the brane configuration in Figure 7, 
$\varphi$ corresponds to 
fundamental strings connecting the $n$ upper flavor D4-branes and
$(\widetilde{N}_c-l)$
color D4-branes(and their mirrors).
The $\Phi_{2n}$ and $ \varphi \varphi$
are $2n \times 2n$ matrices.
The supersymmetric ground state corresponds to
the vacuum expectation values by
$h\Phi_{2n}=
\left[\frac{\mu^2}{\mu_{\phi}(k+1)}\right]^{\frac{1}{2k+1}}  
{\bf 1}_{n} \otimes i \sigma_2$ and $ \varphi =0$ 
\cite{Ahn08-1two}.

\begin{figure}[ht]
   \epsfxsize=2.5in 
\centerline{\epsffile{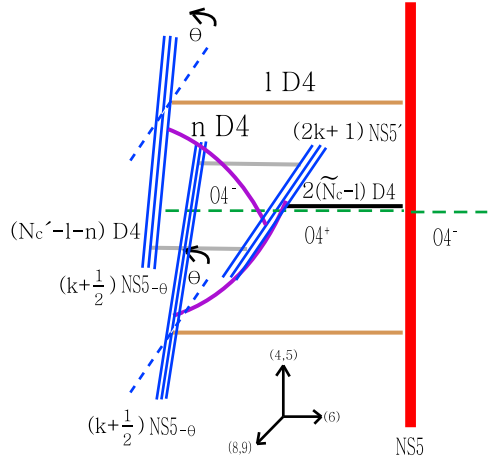}}
   \caption[FIG. \arabic{figure}.]{ 
The nonsupersymmetric meta-brane configuration 
corresponding to Figure 6 when we consider the gravitational potential
of NS5-brane and is obtained 
by moving $n$ flavor D4-branes 
from $(N_c'-l)$ flavor D4-branes of Figure 6.  
The nonzero positive 
$w$ coordinate for $n$ curved flavor D4-branes can be
determined.}
\end{figure}

The full one loop potential for 
$\Phi_{2n}$ and $\varphi$
from (\ref{Dualnew}) and (\ref{Eigennew}) including the one loop result
\cite{ISS} takes the form
\bea
\frac{V}{|h|^2}  =  
|\Phi_{2n}  \varphi|^2   
  +  
| \varphi \varphi-\mu^2 {\bf 1}_{2n} + 
(k+1)h^{2k+1} 
\mu_{\phi}  \Phi_{2n}^{2k+1}|^2 + b |h \mu|^2 \tr \Phi_{2n} \Phi_{2n}, 
\label{pot1}
\eea
where the positive numerical constant $b$ is given by
$b = \frac{(\ln 4-1)}{8\pi^2} (N_c'-N_c-2)$.
Differentiating this potential (\ref{pot1}) with respect to 
$\Phi_{2n}$ and putting $\varphi=0$, one obtains
\bea
\left[- \mu^2 {\bf 1}_{2n}     +(k+1)h^{2k+1} 
 \mu_{\phi} \Phi_{2n}^{2k+1} \right] (2k+1)(k+1) (h^{\ast})^{2k+1}
\mu_{\phi}^{\ast}
\Phi_{2n}^{2k}
 + 2 b
|h \mu|^2 \Phi_{2n} =0.
\nonu
\eea
Sine we assume that $\mu_{\phi}^{\frac{1}{1-2k}} << \mu <<
\Lambda_m$ \cite{GK0710},
the second term of above will be negligible and 
one gets
\bea
  (2k+1)(k+1) \mu^2 (h^{\ast})^{2k+1}
\mu_{\phi}^{\ast}
\Phi_{2n}^{2k}
\simeq 2b
|h \mu|^2 \Phi_{2n}.
\label{cond1}
\eea
Then one arrives at the following form from (\ref{cond1})
\bea
h \Phi_{2n} 
\simeq \left[ \frac{2b }{(2k+1)(k+1) 
\mu_{\phi}} \right]^{\frac{1}{2k-1}}
{\bf 1}_{n} \otimes i \sigma_2,  
M_{2n} \simeq \left[
\frac{\widetilde{N}_c}{\alpha (2k+1)(k+1) 
\Lambda^3} \right]^{\frac{1}{2k-1}} {\bf 1}_{n} 
\otimes i \sigma_2.
\label{vacnew}
\eea
Then the vacuum energy $V$ is given by
$V \simeq 2n h^2 \mu^4$ and
expanding around this solution, one obtains
the eigenvalues for mass matrix for $\varphi$: 
$
m_{\pm}^2
\simeq \left[\frac{2b}{(2k+1)(k+1) 
\mu_{\phi}} \right]^{\frac{2}{2k-1}}  \pm | h \mu|^2$.
For the positive value for these eigenvalues 
one should have
$
 \left[\frac{2b}{(2k+1)(k+1) 
\mu_{\phi}} \right]^{\frac{1}{2k-1}}  >   h \mu
$
which leads to 
the quarks $\varphi$ are massive 
and then the vacuum (\ref{vacnew}) is locally stable
\footnote{The stability of the SUSY breaking vacua 
impies that $h < 
\frac{\mu_{\phi}^{\frac{1}{1-2k}}}{\mu} << 1$. As one flows to IR, the
dimensionless coupling $\mu_{\phi} E^{2k-1}$ at the energy scale $E$
becomes smaller and smaller when $k > 0$. 
At the SUSY breaking scale $E=\sqrt{h}
\mu$, this dimensionless coupling becomes 
$h^{\frac{1-2k}{2}} \mu_{\phi} (h \mu)^{2k-1}$. When $k > 0$, the first
factor $h^{\frac{1-2k}{2}}$ is large. Then the second factor 
should behave as 
$\mu_{\phi} (h \mu)^{2k-1} << 1$ for the higher order deformation to
be  small. This is compatible with stability condition. However, 
the mass of $\varphi$ becomes large when $k
> 0$
and are
integrated out. On the other hand, 
in the quartic case($ k=0$), the qualitative difference arises 
since the first factor $h^{\frac{1-2k}{2}}
<< 1$, as long as $\frac{\mu_{\phi}}{h \mu}$ is not too large then 
the above requirement is valid. In
other words, the mass of $\varphi$  remains
small enough compared to the SUSY breaking scale. In the higher order
case($k > 0$), 
the conditions on $\mu_{\phi}$ and $\mu$ push the SUSY breaking
scale too high for the deformation to stay small enough.  }. 
Note that $|h \Phi_{2n}| << \Lambda_m$.

The upper $(N_c'-l-n)$ flavor D4-branes between 
the upper $NS5_{-\theta}$-branes and 
the $NS5_R'$-branes are related to the corresponding positive eigenvalues 
of $h\Phi$ (\ref{Eigennew}) which is equal to 
$  \left[\frac{\mu^2}{\mu_{\phi}(k+1)}\right]^{\frac{1}{2k+1}} 
{\bf 1}_{(N_c'-l-n)} 
\otimes i \sigma_2$.
The intersection point between the 
upper $(N_c'-l-n)$ D4-branes and the $NS5_R'$-branes is given 
by $(v, w)=(0, +v_{NS5_{-\theta}} \cot \theta)$.
The $n$ upper curved flavor D4-branes between 
the $NS5_{-\theta}$-branes and 
the $NS5_R'$-branes are related to the corresponding positive eigenvalues 
(\ref{vacnew}) of $h\Phi_{2n}$.

Let us consider all the 
$NS5_{-\theta}$-branes and $NS5_R'$-branes
are separated each other. 

$\bullet$ Separated $NS5_{-\theta}$-branes and $NS5_R'$-branes

We displace the $k$ upper $NS5_{-\theta}$-branes and upper $NS5_R'$-branes, 
given in Figure 6, in the $+v$ direction respectively
to two $k$ different points denoted by 
$v_{NS5_{-\theta,j}}$ and $v_{NS5_{R,j}'}$
where $j=1,2, \cdots, k$(and their mirrors in the $-v$ direction).
For convenience, let us put the ``half'' 
$NS5_{-\theta}$-brane as the lowest value of $v$ coordinate
and one remaining $NS5_R'$-brane 
at $v=0$. Then the $k$ upper $NS5_{-\theta}$-branes are located above 
 ``half'' 
$NS5_{-\theta}$-brane and 
$k$ upper $NS5_R'$-branes are located above 
a ``single''  $NS5_R'$-brane whose $v$ coordinate is zero.
The number of color D4-branes
stretched between $j$-th upper $NS5_{R,j}'$-brane and the NS5-brane
is denoted by $\widetilde{N}_{c,j}$
while the number of flavor D4-branes
stretched between the  $j$-th upper $NS5_{-\theta,j}$-brane and  the 
$j$-th upper $NS5_{R,j}'$-brane
is denoted by $N_{c,j}'$. Let us denote 
 $NS5_{-\theta,j=0}$-brane by ``half'' $NS5_{-\theta}$-brane
and  $NS5_{R,j=0}'$-brane by a ``single''  $NS5_R'$-brane.
Then the supersymmetric configurations, labeled by 
sets of nonnegative integers $(\widetilde{N}_{c,0}, \widetilde{N}_{c,1},
\widetilde{N}_{c,2},
\cdots, \widetilde{N}_{c,k})$ and 
sets of nonnegative integers $(N_{c,0}', N_{c,1}',
N_{c,2}',
\cdots, N_{c,k}')$ with 
\bea
\sum_{j=0}^k \widetilde{N}_{c,j}(\equiv N_{c,j}'-N_{c,j}-2)
= \widetilde{N}_c, \qquad
\sum_{j=0}^k N_{c,j}'= N_c'
\label{conforn1}
\eea
can arise.
When all the upper $NS5_{R,j}'$-branes and 
$NS5_{-\theta,j}$-branes($j=0, 1, 2, \cdots, k$) 
are distinct, the low energy  
physics corresponds to $(k+1)$ decoupled supersymmetric 
gauge theories with gauge groups 
\bea
\prod_{j=0}^{k} Sp( \widetilde{N}_{c,j}) \times SO(2N_{c,j}').
\nonu
\eea 
As we approach the origin of 
parameter space, $v_{NS5_{R,j}'}=0$ and
$v_{NS5_{-\theta,j}}=v_{NS5_{-\theta}}$ for all $j$, 
the full $Sp(\widetilde{N}_c) \times SO(2N_c')$
gauge group is restored.

Then each submeson field $M_j \equiv X_j X_j$  is 
in the representation  
$({\bf 1,{N_{c,j}^{\prime}}(2N_{c,j}^{\prime}-1)})$ under the 
gauge group $Sp(\widetilde{N}_{c,j}) \times SO(2N_{c,j}')$.
Note that
a bifundamental $X_j$ is in the representation 
$({\bf 2N_{c,j}, 2N_{c,j}'})$. 
Generic $(v_{NS5_{-\theta,j}}-v_{NS5_{R,j}'})$, 
which is equal to $2\pi \ell_s^2 m_j$, 
is related to a polynomial superpotential for 
$M_j$. The expression  
$\left[m_j {\bf 1}_{N_{c,j}'} 
-\alpha(k+1) M_j^{2k+1} \right]$(where $j=0, 1, 2, \cdots, k$) 
is the
derivative of superpotential with 
respect to the $M_j$ for $Y_j =0$ and has $k$ distinct minima 
$\left[\frac{m_j}{\alpha(k+1)}\right]^{\frac{1}{2k+1}}$ plus zero
with some multiplicities. 
Here $Y_j$  are
dual fields corresponding to $X_j$.

One can write 
the
derivative of superpotential with 
respect to the full meson $M$ for $Y=0$ as
\bea
W_{dual}'(x) = \sum_{i=0}^{N_c'} s_i x^{2(N_c'-i)} \equiv 
-\alpha (k+1) \prod_{j=0}^{k} x^{2l_j} \left(x^2- 
\left[\frac{ m_j}{\alpha(k+1)}\right]^{\frac{1}{2k+1}} \right)^{N_{c,j}'-l_j}
\label{Wprime1}
\eea
where we put the sum of multiplicities of zero eigenvalues as $l$
appeared in
Figure 6
\bea
\sum_{j=0}^{k} l_j=l, \qquad 0 \leq l_j \leq \widetilde{N}_{c,j}.
\label{ljcon1}
\eea 
The order of the polynomial (\ref{Wprime1}) is $2N_c'$.
The integers  $(N_{c,0}',
N_{c,1}',
\cdots, N_{c,k}')$ are the number of the nonzero- plus
zero-eigenvalues 
of the meson
$M$ residing in the different minima.
Thus the set of  
\bea
(N_{c,0}',
N_{c,1}',
\cdots, N_{c,k}'), \qquad (l_0, \l_1, \cdots,
l_k), \qquad \mbox{and} \qquad
(m_0, m_1, \cdots, m_k)
\nonu
\eea
determines the expectation value of $M$ completely. 

Since the eigenvalues for the submeson field $M_j$
are either $0$ or
$\left[\frac{m_j}{\alpha(k+1)}\right]^{\frac{1}{2k+1}}$, 
one takes
$2N_{c,j}' \times 2N_{c,j}'$ matrix $M_j$ with $2l_j$'s 
eigenvalues $0$ and $2(N_{c,j}'-l_j)$'s
eigenvalues $\left[\frac{m_j}{\alpha(k+1)}\right]^{\frac{1}{2k+1}}$ as follows:
\bea
M = \left[\frac{1}{\alpha(k+1)}\right]^{\frac{1}{2k+1}} \left(
\begin{array}{ccccc}
0{\bf 1}_{2l} & 0 & 0 & \cdots  & 0 \\
0 & m_0^{\frac{1}{2k+1}} 
{\bf 1}_{N_{c,0}'-l_0} \otimes i \sigma_2 & 0 & \cdots & 0  \\
\cdot & \cdot & \cdot & \cdots & 0 \\
0 & 0 &  0 & 0 & m_k^{\frac{1}{2k+1}} 
{\bf 1}_{N_{c,k}'-l_{k}}  \otimes i \sigma_2
\end{array}
\right)
\label{M01new1}
\eea
with (\ref{ljcon1}) and (\ref{conforn1}).
In the brane configuration of Figure 8 where we ignore the splitting
of $n_j$ flavor D4-branes for the time being, the $l_j$ of the
upper 
$N_{c,j}'$-flavor D4-branes are connected with $l_j$ of upper 
$\widetilde{N}_{c,j}$-color
D4-branes
and the resulting $l_j$ D4-branes 
stretch from the upper $NS5_{-\theta,j}$-brane to
the NS5-brane directly 
and the intersection point between the 
$l_j$ D4-branes and the NS5-brane is $(v, w)=(+v_{NS5_{-\theta,j}}, 0)$.
This corresponds to  exactly the $l_j$'s eigenvalues $0$ of 
$M_j$ in (\ref{M01new1}).
Now the remaining upper $(N_{c,j}'-l_j)$-flavor D4-branes between 
the upper $NS5_{-\theta,j}$-brane and 
the upper $NS5_{R,j}'$-brane correspond to the remaining eigenvalues 
of $M_j$ in (\ref{M01new1}), i.e.,   
$\left[\frac{m_j}{\alpha(k+1)}\right]^{\frac{1}{2k+1}} 
{\bf 1}_{N_{c,j}'-l_j}$.
The intersection point between the 
upper $(N_{c,j}'-l_j)$ D4-branes and the upper $NS5_{R,j}'$-brane is given 
by $(v, w)=(0, +v_{NS5_{-\theta,j}} \cot \theta)$.

\begin{figure}[ht]
   \epsfxsize=2.5in 
\centerline{\epsffile{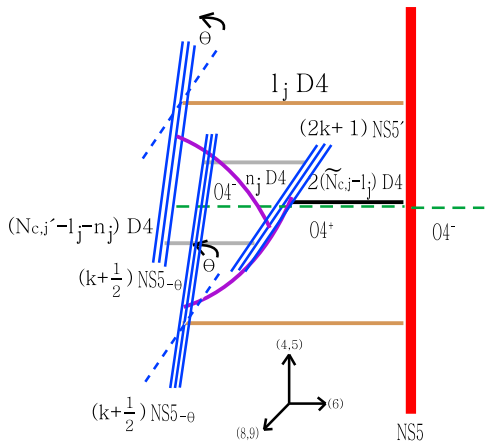}}
   \caption[FIG. \arabic{figure}.]{ 
The nonsupersymmetric meta-brane configuration 
corresponding to Figure 5 when the $NS5_{-\theta}$-branes 
and the $NS5_{R}'$-branes are separated. 
We only draw $(N_{c,j}'-l_j-n_j)$-flavor D4-branes, $n_j$-flavor
D4-branes, 
connecting between 
$NS5_{-\theta,j}$-brane and 
$NS5_{R,j}'$-brane,
$l_j$ D4-branes connecting between 
the $NS5_{-\theta,j}$-brane and
the NS5-brane,
and 
$(\widetilde{N}_{c,j}-l_j)$-color D4-branes connecting between  
the $NS5_{R,j}'$-brane and the NS5-brane. For simplicity, we draw here the
case where $v_{NS5_{R,j}'}$ goes to zero. In general, the $(2k+1)$
$NS5_{R}'$-branes are separated, like 
as $NS5_{-\theta}$-branes, along $v$ direction. }
\end{figure}

By substituting the above $M$ (\ref{M01new1}) into the F-term equation, 
one also obtains
\bea
Y Y = \left(
\begin{array}{cccccc}
-m_1 \Lambda_0 {\bf 1}_{2l_0} & 0 & 0 & \cdots & 0  & 0  \\
\cdot  & \cdot & \cdot & \cdots & \cdot  & 0 \\
0 & 0 & 0 & \cdots & -m_k \Lambda_k {\bf 1}_{2l_k}  & 0 \\ 
0 & 0 & 0 & 0 & 0 & 0{\bf 1}_{2N_c'-2l} 
\end{array}
\right).
\label{solqq1new1}
\eea
In the $l_j$-th vacuum the gauge symmetry is broken to $Sp(\widetilde{N}_{c,j}-l_j)$
and 
the supersymmetric vacuum drawn in Figure 8 with $l_j=0$ has 
$Y_j=0$
and the gauge group 
$Sp(\widetilde{N}_{c,j})$ is unbroken. 

So far, the ground states are supersymmetric. Moreover,
the theory has many nonsupersymmetric meta-stable ground states.
For the IR free region, 
$N_{c,j}+2 < N_{c,j}' < \frac{3}{2}(N_{c,j}+1)$ \cite{ISS,Ahn06-1}, 
the magnetic theory is the effective low energy description of the
asymptotically free electric gauge theory.
When we rescale the submeson field as
$M_{j} = h \Lambda \Phi_{j} $,
then the Kahler potential for $\Phi_{j}$ is canonical and the magnetic
quarks $Y_j$
are canonical near the origin of field space.
Then the magnetic superpotential (\ref{supernew}) 
can be rewritten in terms of $\Phi_j$ and $Y_j$
\bea
W_{dual} = \sum_{j=0}^{k} \left[
 h \Phi_j  Y_j  Y_j 
 +  
\frac{ \mu_{\phi}}{2} h^{2k+2} \tr 
\Phi_j^{2k+2}- h \tr \mu_j^2  \Phi_j \right]
\label{dDualnew}
\eea
with
$
\mu_j^2 = -m_j \Lambda_j$ and 
$\mu_{\phi} = -\alpha \Lambda^{2k+2}$ as before. 
In the brane configuration, this is equivalent to 
deform the Figure 7. Then the $n$ curved flavor D4-branes also are
splitted as $k$ different $n_j$'s(and their mirrors).
One splits 
the $2(N_{c,j}'-l_j) \times 2(N_{c,j}'-l_j)$
block  at the lower right corner of $h\Phi_j$ (\ref{M01new1}) and $Y_j
Y_j$ (\ref{solqq1new1}) 
into blocks of 
size $2n_j$ and $2(N_{c,j}'-l_j-n_j)$ for all $j$ as follows \cite{GK0710}:
\bea
h \Phi & = & \left[\frac{1}{\mu_{\phi}(k+1)}\right]^{\frac{1}{2k+1}} \left(
\begin{array}{ccccc}
0 {\bf 1}_{2l+2n} & 0 & 0 & \cdots  & 0 \\
0 & \mu_0^{\frac{2}{2k+1}} 
{\bf 1}_{N_{c,0}'-l_0-n_0} \otimes i \sigma_2  & 0 & \cdots & 0  \\
\cdot & \cdot & \cdot & \cdots & 0 \\
0 & 0 &  0 & 0 & \mu_k^{\frac{2}{2k+1}} 
{\bf 1}_{N_{c,k}'-l_{k}-n_k}  \otimes i \sigma_2
\end{array}
\right) \nonu \\
& + & \mbox{diag}(
0 {\bf 1}_{2l}, h \Phi_{2n_0}, \cdots, h \Phi_{2n_k}, 
0 {\bf 1}_{2N_c'-2l-2n})
\label{M02new}
\eea
and
\bea
Y  Y  & = & \left(
\begin{array}{cccccc}
\mu_1^2  {\bf 1}_{2l_0} & 0 & 0 & \cdots & 0  & 0  \\
\cdot  & \cdot & \cdot & \cdots & \cdot  & 0 \\
0 & 0 & 0 & \cdots & \mu_{k}^2  {\bf 1}_{2l_k}  & 0 \\ 
0 & 0 & 0 & 0 & 0 & 0 {\bf 1}_{2N_c'-2l} 
\end{array}
\right) \nonu \\
& + &  \mbox{diag}(
0{\bf 1}_{2l}, \varphi_{2n_0} \varphi_{2n_0}, \cdots,
\varphi_{2n_k} \varphi_{2n_k}, 
0 {\bf 1}_{2N_c'-2l-2n}).
\label{solqq2new}
\eea
Here 
$\varphi_{2n_j}$ are 
$2n_j \times 2(\widetilde{N}_{c,j}-l_j)$
matrices and correspond to $2n_j$-flavors of fundamentals of
the gauge group $Sp(\widetilde{N}_{c,j}-l_j)$ which is unbroken.
Let us denote the sum of $n_j$ by $n$ appeared in Figure 7
\bea
\sum_{j=0}^k n_j = n, \qquad 0 \leq n_j \leq N_{c,j}'-l_j.
\nonu
\eea
In the brane configuration from Figure 8, 
they correspond to 
fundamental strings connecting between the $2n_j$-flavor D4-branes and
$2(\widetilde{N}_{c,j}-l_j)$-color D4-branes. Moreover,
the $\Phi_{2n_j}$ and ${ \varphi}_{2n_j} 
\varphi_{2n_j}$
are $2n_j \times 2n_j$ matrices.
The supersymmetric ground state corresponds to the vacuum expectation
values by
$h\Phi_{2n_j}=\left[\frac{\mu_j^2}{\mu_{\phi}(k+1)}\right]^{\frac{1}{2k+1}} 
{\bf 1}_{n_j} \otimes i \sigma_2 $ and $
\varphi_{2n_j}=0$.
The $\Phi_j$ which is a $2N_{c,j}' \times 2N_{c,j}'$ 
contains a submatrix $\Phi_{2n_j}$.

The full one loop potential for 
$\Phi_{2n_j}$ and $\varphi_{2n_j}$
from (\ref{dDualnew}), (\ref{M02new}) and (\ref{solqq2new}) takes the form
\bea
\frac{V}{|h|^2}  =  \sum_{j=1}^{k} \left[
|\Phi_{2n_j}  \varphi_{2n_j}|^2   
  +  
| \varphi_{2n_j} \varphi_{2n_j}-\mu_{j}^2 {\bf 1}_{2n_{j}} + 
(k+1)h^{2k+1} 
\mu_{\phi}  \Phi_{2n_j}^{2k+1}|^2 + b_j |h \mu_{j}|^2 \tr \Phi_{2n_j}
\Phi_{2n_j} 
\right], 
\nonu
\eea
where
$b_j = \frac{(\ln 4-1)}{8\pi^2} (N_{c,j}'-N_{c,j}-2)$ for all $j$ 
\cite{ISS}.
Differentiating this potential with respect to 
$\Phi_{2n_j}$ and putting $\varphi_{2n_j}=0$, one obtains
\bea
\left[- \mu_j^2 {\bf 1}_{2n_j}     +(k+1)h^{2k+1} 
 \mu_{\phi} \Phi_{2n_j}^{2k+1} \right] (2k+1)(k+1) (h^{\ast})^{2k+1}
\mu_{\phi}^{\ast}
\Phi_{2n_j}^{2k}
 + 2b_j
|h \mu_j|^2 \Phi_{2n_j} =0.
\nonu
\eea
Sine the higher order term superpotential plays the role of 
small perturbation and we assume that $\mu_{\phi}^{\frac{1}{1-2k}} << \mu_j <<
\Lambda_{m,j}$ \cite{GK0710},
the second term of above will be negligible and 
one gets
\bea
  (2k+1)(k+1) \mu_j^2 (h^{\ast})^{2k+1}
\mu_{\phi}^{\ast}
\Phi_{2n_j}^{2k}
\simeq 2b_j
|h \mu_j|^2 \Phi_{2n_j}.
\nonu
\eea
In order to see the structure of this solution, 
we consider the particular case where 
all the parameters are real.
The general case is, in principle, straightforward.
Then one arrives at the following form
\bea
h \Phi_{2n_j} 
& \simeq & \left[ \frac{2b_j }{(2k+1)(k+1) \mu_{\phi}} \right]^{\frac{1}{2k-1}}
{\bf 1}_{n_j} \otimes i \sigma_2, \qquad \mbox{or} \nonu \\  
M_{2n_j} & \simeq & \left[
\frac{\widetilde{N}_{c,j}}{\alpha (2k+1)(k+1) 
\Lambda^3} \right]^{\frac{1}{2k-1}} {\bf 1}_{n_j} \otimes i \sigma_2.
\label{vac1new1}
\eea
Then the vacuum energy $V$ is given by
$V \simeq \sum_{j=0}^k 2n_j h^2 \mu_j^4$ and
expanding around this solution, one obtains
the eigenvalues for mass matrix for $\varphi_{2n_j}$: 
$
m_{\pm}^2
\simeq \left[\frac{2b_j}{(2k+1)(k+1) 
\mu_{\phi}} \right]^{\frac{2}{2k-1}}  \pm | h \mu_j|^2$.
For the positive value for these eigenvalues, 
one should have
$
 \left[\frac{2b_j}{(2k+1)(k+1) 
\mu_{\phi}} \right]^{\frac{1}{2k-1}}  >   h \mu_j
$
which leads to 
the quarks $\varphi_{2n_j}$  are massive 
and then the vacuum (\ref{vac1new1}) is locally stable. 

The $(N_{c,j}'-l_j-n_j)$ flavor D4-branes between 
the $NS5_{-\theta,j}$-brane and 
the $NS5_{R,j}'$-brane are related to the corresponding eigenvalues 
of $h\Phi$ (\ref{M02new}), i.e.,  
$ 
\left[\frac{\mu_j^2}{\mu_{\phi}(k+1)}\right]^{\frac{1}{2k+1}} 
{\bf 1}_{N_{c,j}'-l_j-n_j} \otimes i \sigma_2$ and 
the intersection point between the 
$(N_{c,j}'-l_j-n_j)$ D4-branes and the $NS5_{R,j}'$-brane is also given 
by $(v, w)=(0, +v_{NS5_{-\theta,j}} \cot \theta)$.
Moreover, 
the remnant $2n_j$ flavor D4-branes between 
the $NS5_{R,j}'$-brane and the $NS5_{-\theta,j}$-brane 
are related to the corresponding eigenvalues 
(\ref{vac1new1}) 
of $h\Phi_{2n_j}$. 

Then, the meta-stable states, for fixed $k$ and $\theta$, 
are classified by the number of various D4-branes and the positions of
multiple NS-branes. That is, the former is given by 
$i)$ $N_{c,j}$ which appears in the number of dual color D4-branes
and  in $h \Phi_{2n_j}$ (\ref{vac1new1}),  
$ii)$ $N_{c,j}'$ which is encoded also in 
the number of dual color D4-branes, 
in ``straight'' flavor D4-branes, in the
multiplicities of expectation value $h \Phi$ (\ref{M02new}), and  in
$h \Phi_{2n_j}$, 
$iii)$ $l_j$
which characterizes the number of splitting D4-branes
and the
multiplicities of 
expectation value $h\Phi$, and 
$iv)$ $n_j$ which is
the number of ``curved'' flavor D4-branes and determines the
multiplicities of the expectation value 
for $h \Phi_{2n_j}$  and the
multiplicities of expectation value $h\Phi$. 
The latter is given by 
$i)$ $v_{NS5_{-\theta,j}}$ which provides the locations of $l_j$ D4-branes
and stright flavor D4-branes, and 
$ii)$ $v_{NS5_{R,j}'}$ which determines the locations of dual 
color D4-branes and  straight flavor  
D4-branes. 

By applying the Seiberg dual to the $SO(2N_c')$ gauge group, 
one can construct the magnetic brane configuration.
Then the dual gauge group is given by
$
Sp(N_c) 
\times SO(2\widetilde{N}_c'=2N_c-2N_c'+4)$ and
the matter contents are the field $Y$ in the  
representation $({\bf 2N_c, 2\widetilde{N}_c'})$ 
and  the gauge-singlet $M$ is
in the representation
$({\bf N_c(2N_c+1), 1})$ under the dual gauge group \cite{Ahn08-1two}.
The discussion for the supersymmetric or nonsupersymmetric 
vacua in previous description 
can be applied here also without any difficulty.
Or if we change the charges of orientifold 4-plane in Figures 5-8,
then the symplectic(orthogonal) gauge group 
changes to the orthogonal(symplectic) gauge group and the matter
contents also change.  One can also analyze the (meta-)stable 
states.

\section{Conclusions and outlook }

We have found the type IIA
nonsupersymmetric meta-stable brane
configuration, presented by Figures 3 and 4, 
consisting of $(2k+1)$ NS-branes and different kinds of 
D4-branes 
where the electric gauge theory superpotential 
(\ref{superelectric}) initially has 
an order $2(k+1)$ polynomial for the bifundamentals.
There exists a rich pattern of nonsupersymmetric meta-stable states in
the presence of $n_j$-flavor D4-branes as well
as the supersymmetric stable ones, in Figures 2 and 4, 
without those $n_j$-flavor D4-branes.
By adding the orientifold 4-plane to this brane
configuration,
we have also described the intersecting brane configuration of type IIA 
string theory, presented by Figures 7 and 8, consisting of $(4k+3)$
NS-branes, 
various different kinds of 
D4-branes, and an orientifold 4-plane 
where the electric gauge theory superpotential 
(\ref{superelectric1}) has 
an order $4(k+1)$ polynomial for the bifundamentals.
The presence of orientifold 4-plane determines the gauge groups,
matter contents and the positions of NS-branes and D4-branes. 

It would be interesting to see whether 
the quartic case for the supersymmetric gauge theory with adjoint
matter \cite{Ahn06} 
can be generalized to the higher order superpotential by increasing
the number of NS5-branes, like as this
paper.
One can also study, in principle, how the meta-stable brane configurations
for higher order superpotential case occur 
when there exists an orientifold 6-plane. For example,
in the brane configurations 
\cite{Ahn0701,Ahn0702,Ahn0705,Ahn0707,Ahn0708,Ahn0710,Ahn0711}, it is
natural to ask what happens if we increase the number of NS5-branes by
looking at the supersymmetric magnetic brane configurations and
analyzing the brane motion for the flavor and color D4-branes.

\vspace{.7cm}

\centerline{\bf Acknowledgments}

This work was supported by grant No.
R01-2006-000-10965-0 from the Basic Research Program of the Korea
Science \& Engineering Foundation.  
I would like to thank KIAS(Korea Institute for 
Advanced Study) for hospitality  where
this work was undertaken.

\end{document}